\documentclass[%
reprint,
amsmath,amssymb,
 aps,
 prd,
floatfix,
]{revtex4-2}

\usepackage{graphicx}
\usepackage{dcolumn}
\usepackage{bm}
\usepackage{hyperref}
\usepackage{aas_macros}
\usepackage[font=small,labelfont=bf,
   justification=justified,
   format=plain]{caption}
\usepackage{xcolor}

\begin{document}

\preprint{arXiv:xxxx.xxxx}

\title{Galaxy Deblending using Residual Dense Neural network}

\author{Hong Wang}
 \affiliation{Stony Brook University, Stony Brook, NY 11794}
 \affiliation{Computational Science Initiative, Brookhaven National Laboratory, Upton, NY 11973}
 
    \author{Sreevarsha Sreejith}
 \affiliation{Physics Department, Brookhaven National Laboratory, Upton, NY 11973}

\author{Yuewei Lin}
 \affiliation{Computational Science Initiative, Brookhaven National Laboratory, Upton, NY 11973}

\author{An\v{z}e Slosar}
 \affiliation{Physics Department, Brookhaven National Laboratory, Upton, NY 11973}

\author{Shinjae Yoo}
 \affiliation{Computational Science Initiative, Brookhaven National Laboratory, Upton, NY 11973}

\date{\today}

\begin{abstract}
  We present a new neural network approach for deblending galaxy images in astronomical data using Residual Dense Neural network (RDN) architecture. We train the network on synthetic galaxy images similar to the typical arrangements of field galaxies with a finite point spread function (PSF) and realistic noise levels. The main novelty of our approach is the usage of two distinct neural networks: i) a \emph{deblending network} which isolates a single galaxy postage stamp from the composite and, ii) a \emph{classifier network} which counts the remaining number of galaxies. The deblending proceeds by iteratively peeling one galaxy at a time from the composite until the image contains no further objects as determined by the classifier, or by other stopping criteria. By looking at the consistency in the outputs of the two networks, we can assess the quality of the deblending. We characterize the flux and shape reconstructions in different quality bins and compare our deblender with the industry standard, \texttt{SExtractor}. We also discuss possible future extensions for the project with variable PSFs and noise levels.
\end{abstract}

\maketitle

\section{Introduction}
Astronomical survey science is entering a decade which will witness a surfeit of new optical imaging data that will transform astronomical and cosmological research. The new datasets obtained from upcoming ground-based and space-based observing facilities will image vast areas of the sky at unprecedented depths. Among the most important of these will be the Legacy Survey of Space and Time (LSST) from the Vera Rubin Observatory \cite{0912.0201}, which will complement not only the already existing datasets from Dark Energy Survey (DES, \cite{1708.01530}) and Hyper-Suprime Cam (HSC,\cite{1809.09148}), but also the data from upcoming space-based surveys like Euclid \cite{1001.0061}. Due to the large field of view, most of these surveys will not be able to employ adaptive optics with a point spread function (PSF) that is comparable to the object size (around 0.5-1 arcsecs on the ground and 0.1 arcsecs in space). At the depths of these surveys, this will result in blending, affecting over half of the galaxies in the survey (for example, in \cite{2005.12039}, the authors estimate that approximately 60\% of the objects in the HSC are affected by this problem).

Blending, especially if undetected, can introduce serious systematic errors in survey analysis. These include potential catastrophically wrong inference of fluxes, leading to biased photometric redshifts, biased estimates of shear and local object density dependent systematics, that can interact with galaxy and cluster detections. The main approach to solving the blending problem is two-pronged. On one hand, there is the realization that source separation (deblending) will never be perfect and we should focus on understanding its properties and effects on the analysis. This is done through simulations and artificial source injections into real data. On the other hand, the better the deblender is, the smaller the corrections needed to make the analysis unbiased. Therefore, we need both good deblenders and good schemes for understanding their imperfections.

In principle, deblending is a well defined problem. The basic model is that the images of individual galaxies are combined on the projected plane, assuming perfect transparency (i.e. intensities add up), and then observed through the atmosphere and telescope with known PSF and noise properties. The most sophisticated deblenders on the market combine machine learning approaches for setting priors on galaxy shapes, and physical modelling for things that we can model explicitly, like the noise and the PSF \cite{1802.10157,1912.03980}.

Several machine learning based methods have been proposed recently, in order to grapple with the galaxy deblending problem. ~\cite{reiman2019deblending} designed a branched deblender with generative adversarial networks, which can work on the deblending of images with two overlapped galaxies. ~\cite{boucaud2020photometry} developed a framework for measuring the photometry of blended galaxies as well as to perform segmentation with a standard convolutional neural network (CNN) and a U-Net. ~\cite{2005.12039} introduced an algorithm where a Variational Auto Encoder(VAE)-like neural network was used for galaxy deblending. Most of the current generation neural network based galaxy deblenders require the galaxy to be located at the center of the image in order to give the network a sign as to which galaxy to recover, which is often impractical and prevents the model from being used iteratively, i.e., it will degrade the model's performance on the image after the first galaxy is removed. Besides, the number of galaxies in the blended image for the previous methods is always fixed, and to the best of our knowledge, there is no neural network based galaxy deblending framework that can work on images with an arbitrary and unknown number of galaxies. In this paper, we propose an innovative deblending framework with a deblender, and a classifier that can deblend galaxies from a blended image iteratively without any prior information about the number of galaxies. In addition, since the deblender recovers galaxies based on their luminosity, our framework has no constraint on the position of the galaxies. Nevertheless, like other machine learning approaches to deblending, our work remains exploratory as we aim to better understand the applicability of neural networks to astronomical image analysis.

This paper is organized as follows: in Section~\ref{sec:NN_Training}, we introduce the architecture of our framework including the deblender and the classifier, and the experimental settings used to train the model. In Section~\ref{sec:Result}, we present the experimental results and the comparison with the industry standard deblending method - Source Extractor (SExtractor, \cite{B&A1996}). The discussion and conclusion follow, in the last Section~\ref{sec:Conclusion}.

\section{Neural Network Architecture and Training} \label{sec:NN_Training}

\newcommand{\deblender}{\emph{deblender}}
\newcommand{\classifier}{\emph{classifier}}

\subsection{Proposed Framework} \label{fwdesc}
Our goal is to deblend galaxies from astronomical images with an arbitrary and unknown number of overlapped galaxies. The proposed framework consists of two components: a \deblender, which isolates the image corresponding to a single galaxy from an astronomical image, and a \classifier, which counts how many galaxies remain in the image. The \deblender~and \classifier\ are then used iteratively to separate the scene into its constituent galaxy images. This is illustrated in the FIG. \ref{fig:framework} and represented in the following meta-code:
\begin{verbatim}
while True:
    num_galaxies = classifier(image)
    if num_galaxies == 0:
        exit loop
    deblended_galaxy = deblender(image)
    image -= deblended_galaxy
\end{verbatim}

\begin{figure}[htbp]
  \centering
  \includegraphics[width=\linewidth]{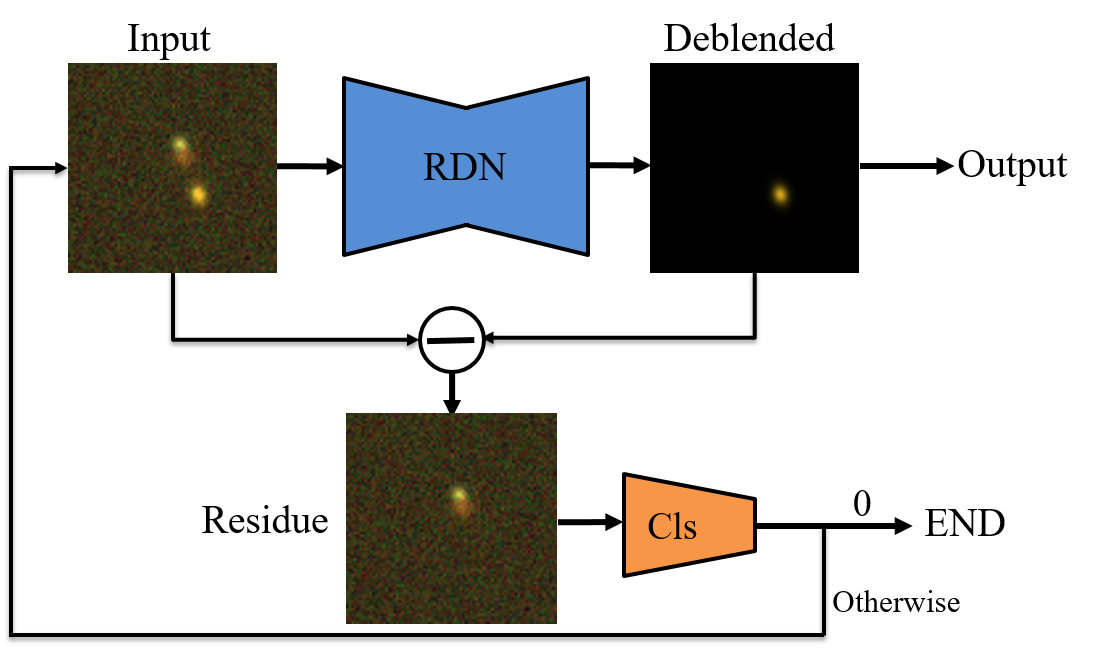}
  \caption{Figure showing the proposed deblender-classifier framework represented as meta-code in Section \ref{fwdesc}. }
  \label{fig:framework}
\end{figure}
Specifically, given a noisy blended image with multiple galaxies, the deblender will take it as the input and output a noiseless image with a single galaxy as it is trained. This single-galaxy image is then subtracted from the input of the deblender to get the residual image, which is another noisy blended image. The classifier is used to determine if there are further galaxies in the residual image, and if there are, this process is repeated until there is no galaxy left in the residual image. 

In an ideal case, the classifier will detect one fewer galaxy at each step and the process will stop when there are no more galaxies left. We call such deblends \emph{High quality} deblends. In a non-ideal case, one of the two following scenarios often play out. The number of galaxies in the image does not decrease by one galaxy per iteration, but the process still eventually comes to a halt with zero galaxies detected in the final image. When this happens, we refer to the result as \emph{Medium quality} deblends. The third case is if the process gets stuck in an infinite loop where the \classifier\  maintains that there are more than zero galaxies in the image, but the deblender fails to locate them. Then, if the classifier predicts the same non-zero number of galaxies in the residual image for three consecutive iterations we terminate the iteration and call the results \emph{Low quality} deblends.

\begin{figure}[htbp]
    \centering
    \includegraphics[width=\linewidth]{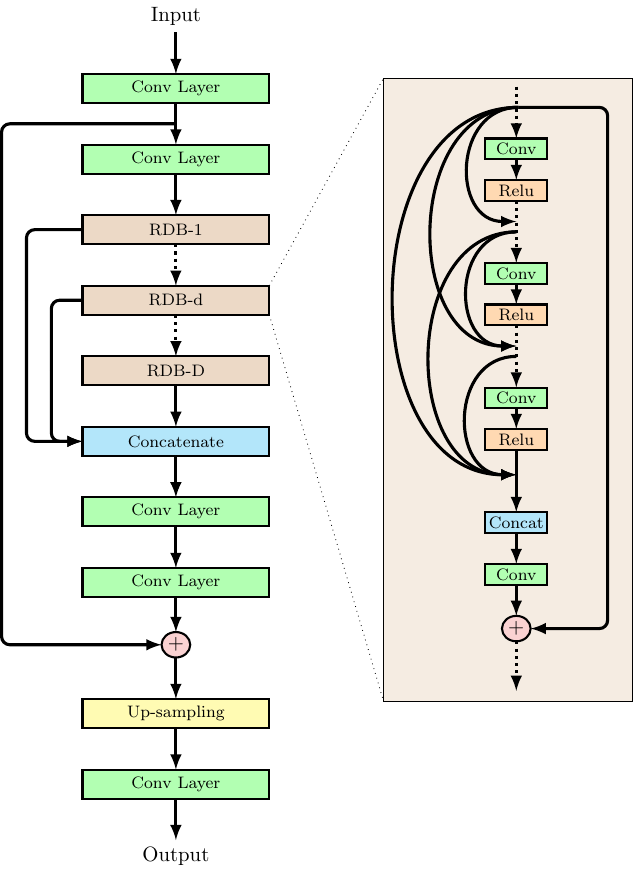}
    \caption{Figure showing the architecture of the RDN used as the deblender in this work. }
    \label{fig:rdn}
\end{figure}

\subsection{Deblender}
Given an astronomical image with multiple overlapped galaxies, the deblender aims to deblend one galaxy from it. A Residual Dense Network (RDN)~\cite{zhang2018residual} is trained as the deblender in this framework. RDN shows superior performance in image super-resolution~\cite{zhang2018residual} and image restoration~\cite{zhang2020residual}. For the deblending task, the RDN will take the noisy blended images with multiple galaxies as input and will give a noiseless image containing the brightest galaxy as the output.

FIG.~\ref{fig:rdn} shows the architecture of the deblender in our framework. It starts with a shallow feature extractor, which includes two convolutional layers. Following this is the main component of the RDN, namely the residual dense blocks (RDBs). Each RDB contains $C$ convolutional layers with a ReLU activation function. The layers in the RDB are densely connected to make better use of the local features. The deblender contains $D$ RDBs in total. An advantage of the original RDN is the design of feature fusion and residual learning in multiple levels, i.e. locally within each RDB and globally among different RDBs, which will extract hierarchical features and provide the full usage of feature information. For the galaxy deblending problem in which the input images are likely to be overlapped blended galaxies with noise, the existence of those direct and dense connections for the features will help to preserve the detailed morphology and color information of the galaxies during deblending and will lead to recovering better, the pertinent galaxy. The efficient sub-pixel convolutional neural network (ESPCN) followed by a convolutional layer forms the up-sampling net in the RDN. Since the model in our framework is used for deblending, the output image will have the same size as the input image. Therefore, the scale factor in the deblender is set to be $1$. During training, the deblender will take the noisy blended images as the input and output noiseless images with a single galaxy. In order to avoid confusing the deblender, we train the model to output the brightest galaxy in the input image all the time. $\ell_1$-norm is used to compute the difference between the prediction of the deblender and the ground truth, thus the loss function for the deblender is written as:

\begin{equation}
    l^\theta_{\rm RDN}(I, I_{gt}) = \frac{1}{WH}\sum_{x=1}^{W}\sum_{y=1}^{H} \left|f(I, \theta)_{x,y}-(I_{gt})_{x,y} \right| \label{eq:RDN}
\end{equation}

where $I$ is the input noisy blended image and $f_{\theta}(I)$ is the output of the RDN with $\theta$ referring to the parameters in the model. $I_{GT}$ denotes the ground truth image, which is a noiseless image containing the brightest galaxy. 

\subsection{Classifier}
For the classifier portion of the deblending network, we use a VGG-16 network~\cite{simonyan2014very}. The VGG-16 contains 13 convolutional layers and 3 fully connected layers. In our framework, it is modified to give four classes as output, i.e., the classifier can tell if the images have 0, 1, 2 or 3 galaxies.

In the first phase of training, the classifier is trained on images with galaxies ranging from 0 to 3. During the second phase in the training process where the deblender and the classifier are trained jointly, the deblended images from the deblender will also be passed through the classifier to see if the RDN has deblended only one galaxy from the blended image. This loss will also be used to update the RDN's parameters. Thus, the training set of the classifier contains both noiseless and noisy images with 0 to 3 galaxies. The cross-entropy (CE) loss is utilized to train the classifier parameterized by $\phi$ in Eq.~\ref{eq:CE} where $y$ represents the one-hot ground truth label. 

\begin{equation}
    l^\phi_{CE}(I) = -y^{\rm T}\cdot\log(g(I, \phi)) \label{eq:CE}
\end{equation}

\subsection{Training \& Test Data}

The galaxy images (blended and otherwise) used in this work has been generated using BlendingToolKit \footnote{\url{https://github.com/LSSTDESC/BlendingToolKit}}. It is a complete framework whose functionalities include generation of images with blended objects as well as providing measurements of the deblending and detection performance. It relies on the industry standard \texttt{galsim} \cite{2015A&C....10..121R} packages to make the actual renderings of the galaxies and is built on top of the \texttt{WeakLensingDeblending} \footnote{\url{https://github.com/LSSTDESC/WeakLensingDeblending}} package.

We first generate noiseless images of single galaxies, which can be considered as the pre-blended ground truth images. We employ the LSST DM galaxy catalog with a span of 1 square deg (\texttt{OneDegSq.fits} supplied with the BlendingToolKit). This catalog is derived from the \texttt{CatSim} galaxy catalog used by the LSST Data Management. It is based on dark matter haloes from the Millennium simulation and a semi-analytic baryon model from \cite{2006MNRAS.366..499D}. The semi-analytic model was calibrated in luminosity, color and morphology on low redshift galaxies. LSST cosmological catalogs were generated from this model by constructing a lightcone, covering redshifts $0<z<6$ by tiling $58$ simulation snapshots with halo masses down to $2.5\times 10^9 M_\odot$. The final catalog comprises of a $4.5 \times 4.5$ degree footprint on the sky, of which one square degree is sampled in \texttt{OneDegSq.fits}. The resulting galaxies are thus somewhat realistic in morphologies, luminosities and colours, but lack the the full morphological richness of real galaxies. 

We use the default options, which include pixel scale of 0.2 arcsec/pixel and impose a magnitude cut on the i-band, $i<25.3$.  This cut corresponds to the LSST gold sample cut and thus represents a set of galaxies that should be measured with reasonable accuracy by the LSST. While it is true that fainter interlopers might negatively affect the deblending procedure, we have ignored this in this exploratory paper. The default option also includes a constant Gaussian PSF with a FWHM of $0.73$, $0.70$ and $0.67$ arcsec in the $g$, $r$ \& $i$ bands respectively. Then, the noiseless blended images are generated by a pixelwise summation between these single galaxy images. Additional random Poisson noise is added to the image to get the noisy blended images corresponding to the 10-year depth survey depth of the LSST. We selected $g$, $r$ and $i$ bands from the resultant images and converted them into \texttt{RGB} images. The original dimension of the generated images is ($120$, $120$, $3$). We crop each image to ($80$, $80$, $3$) from the center. The pixel values of the blended images will be normalized to $[0, 1]$ before entering the framework and the pre-blended images will be scaled to $[-1, 1]$ following~\cite{reiman2019deblending} . 

Galaxies in the current draft are randomly positioned (with appropriate buffer around the edges of the image), corresponding to the random chance alignments between galaxies, which is cause of the majority of blends. Blending occurring due to physical associations of galaxies would in any case require a different training set. 

Unlike the other machine learning based deblending methods where one of the galaxies has to be located at the center of the image~\cite{reiman2019deblending, boucaud2020photometry}, there is no constraint on the positions of the galaxies for our deblender, because we train it to output the object with the highest luminosity. This is also the reason why our framework can work iteratively. The use of the pixelwise summation when creating the blended images is also instrumental in making the iterative process possible. During training, the noisy blended images are treated as the input and the noiseless image with the brightest galaxy as the pre-blended ground truth for the deblender. Our goal is to train a deblender that can deblend and denoise at the same time and will always recover the brightest galaxy from the blended images. The classifier is trained with $0-3$ galaxies and for each of the four classes, half of the training data contains noiseless images while the other half contains noisy images. This is essential in order to train a classifier that can estimate the number of galaxies in both noiseless and noisy images at the same time.

We generate $50,000$ $2$-galaxy blended images as the training set for the deblender using BlendingToolKit. The test set contains $1000$ blended images with $2$ galaxies in each. 
There are also several extended test sets for $3$ to $7$-galaxy blended images for generalization, each containing $1000$ images.
For the classifier, we generated blended images containing $0$, $1$, $2$ and $3$ galaxies, where there are $100,000$ images of each class (making it $400,000$ in total) -  half of these are noiseless images and the other half noisy. The test set for the classifier contains $2000$ images of each class and $8000$ images in total. The test set for the classifier also has the same ratio of noiseless and noisy images. The classifier gives $4$ classes as output indicating the number of galaxies in the input image.

\subsection{Training procedure}

The training process can be divided into two phases. In the first phase, the deblender and the classifier are pre-trained separately until they perform reasonably. Then they are fine-tuned jointly to boost the performance of both, while working in the iterative setting. Specifically, we expect that the classifier will force the deblender to output only one galaxy, while the deblender will provide more intermediate images to motivate the classifier to have better discrimination ability.

During the pre-training of the RDN, the noisy $2$-galaxy blend images are used as the input and the noiseless images with the brighter single galaxy as the ground truth. Eq.~\ref{eq:RDN} is the objective function during this process. A model trained in this way will have the ability to not only deblend the brighter galaxy, but also to simultaneously denoise . Meanwhile, in the first training phase the classifier is trained on the dataset containing both noisy and noiseless blended images with $0$, $1$, $2$ and $3$ galaxies. This pre-training phase makes the trained classifier able to classify both noisy and noiseless images and the trained RDN to determine the end of the whole workflow. They will be used together during the second phase of the training.

During the second phase, the deblender and the classifier will be trained jointly. The framework will operate as it is designed, i.e., the deblender will be run twice because the input contains two galaxies. The blended image, denoted as $I_{{\rm blend}}$ will be passed through the deblender to get the deblended $I_{{\rm deblend}-1}$, which is compared with the ground truth $I_{{\rm gt}-1}$ to build the $\ell_1$-norm loss $l_{\rm RDN}(I_{{\rm deblend}-1}, I_{{\rm gt}-1})$. The first residual image $I_{{\rm res}-1}$ is the subtraction between $I_{{\rm blend}}$ and $I_{{\rm deblend}-1}$. Generally, the residual images are fainter than the blended images and therefore they will be normalized by dividing with the maximum pixel values. Then, the deblender will predict and scale back to recover the second galaxy $I_{{\rm deblend}-2}$ from $I_{{\rm res}-1}$ and another residual image $I_{{\rm res}-2}$ will be calculated. For simplicity, we use $\mathbb{S}=\{I_{{\rm blend}}, I_{{\rm deblend}-1}, I_{{\rm deblend}-2}, I_{{\rm res}-1},  I_{{\rm res}-2}, I_{{\rm gt}-1},  I_{{\rm gt}-2}\}$ to denote all the images. In this phase, the classifier will be optimized by all the available images in $\mathbb{S}$ as formulated in Eq.~\ref{eq:VGG-2}. Since $\mathbb{S}$ is class-imbalanced with more $1$-galaxy images than $0$ and $2$-galaxy images, $\alpha_I$ is used as a weight. For $1$-galaxy images, $\alpha_I=0.2$, otherwise $\alpha_I=1$.  For the deblender, in addition to $l_{\rm RDN}(I_{{\rm deblend}-1}, I_{{\rm gt}-1})$, it is updated based on a loss from the classifier's prediction on the deblended images. The objective function for the second phase can be formulated as Eq.~\ref{eq:RDN-2}. $\lambda$ here is a trade-off coefficient.

\begin{equation}
    l^\phi_{phase-2} = \sum_{I \in \mathbb{S}} \alpha_I l_{CE}(I) \label{eq:VGG-2}
\end{equation}

\begin{equation}
    l^\theta_{phase-2} = l_{\rm RDN}(I_{{\rm deblend}-1}, I_{{\rm gt}-1})+\lambda l_{CE}(I_{{\rm deblend}}) \label{eq:RDN-2}
\end{equation}

\begin{figure}[htbp]
    \centering
    \includegraphics[width=\linewidth]{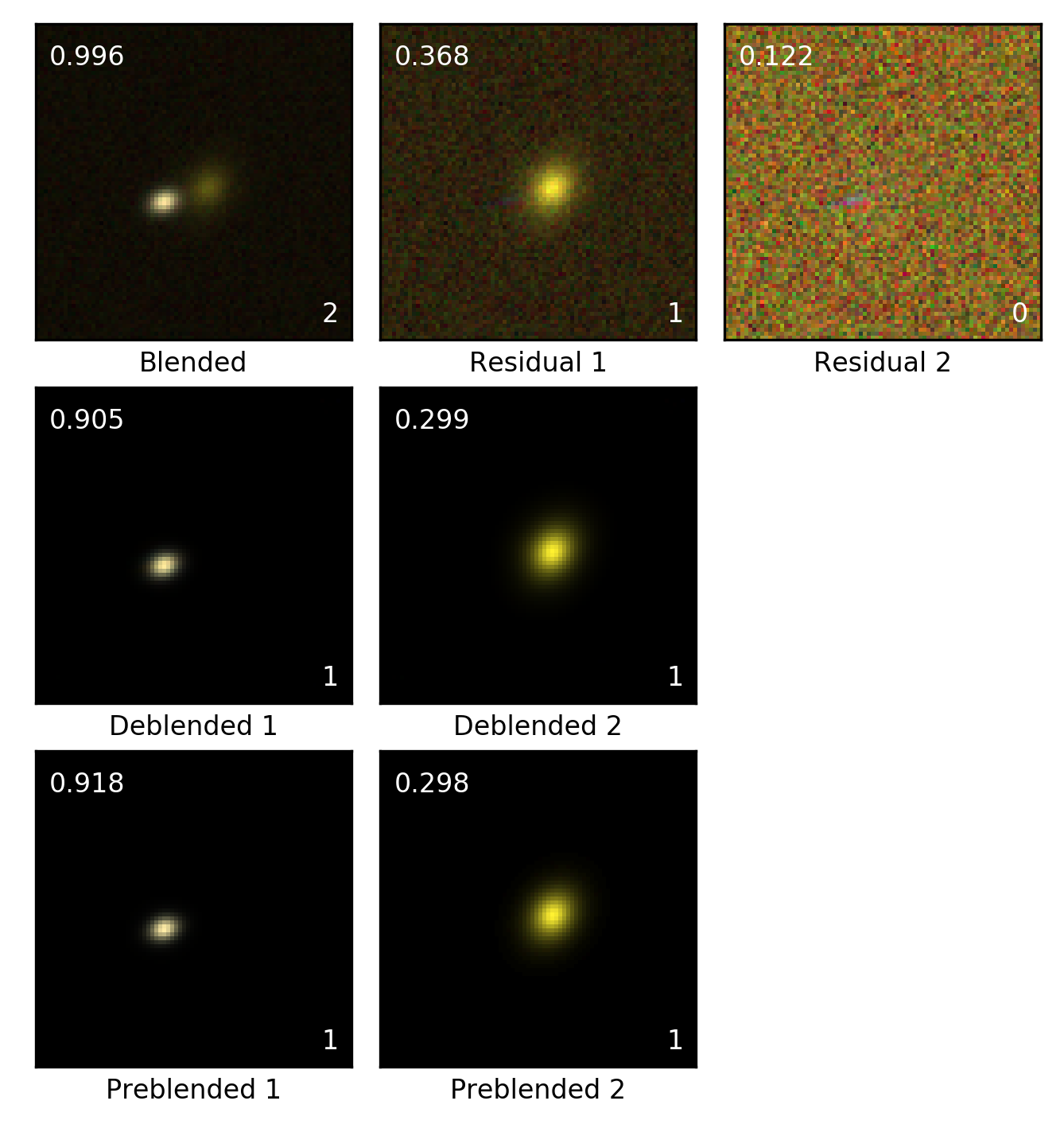}
    \includegraphics[width=\linewidth]{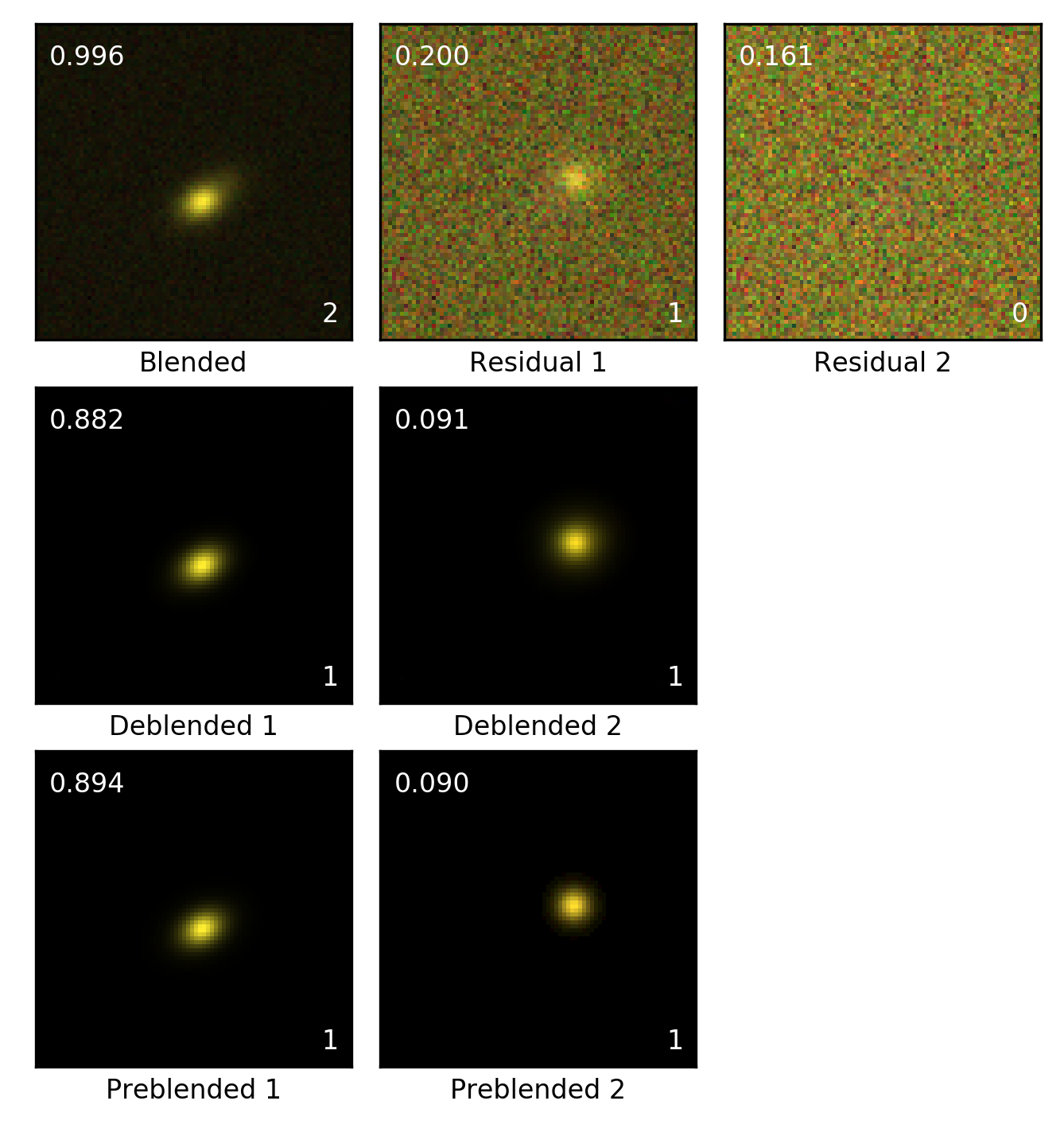}
    \captionsetup{justification=justified}

    \caption{Figure showing iterative results for $2$-galaxy blended images. In each of the two examples shown here, the top row shows the noisy image as it progresses through the iterative process with the input image on the left and the remaining noise image (after the galaxies were subtracted) on the right. The middle row shows the galaxy images that were isolated from the input image by the \deblender. The bottom row shows the ground truth, which can be visually compared with the deblended images above. All the images were rescaled by its maximum pixel value which is indicated at the top left. 
    The other number at the bottom right corner of each image represents the number of remaining galaxies as determined by the classifier.
    }
    \label{fig:iterative_blend2}
\end{figure}

\begin{figure}[htbp]
    \centering
    \includegraphics[width=\linewidth]{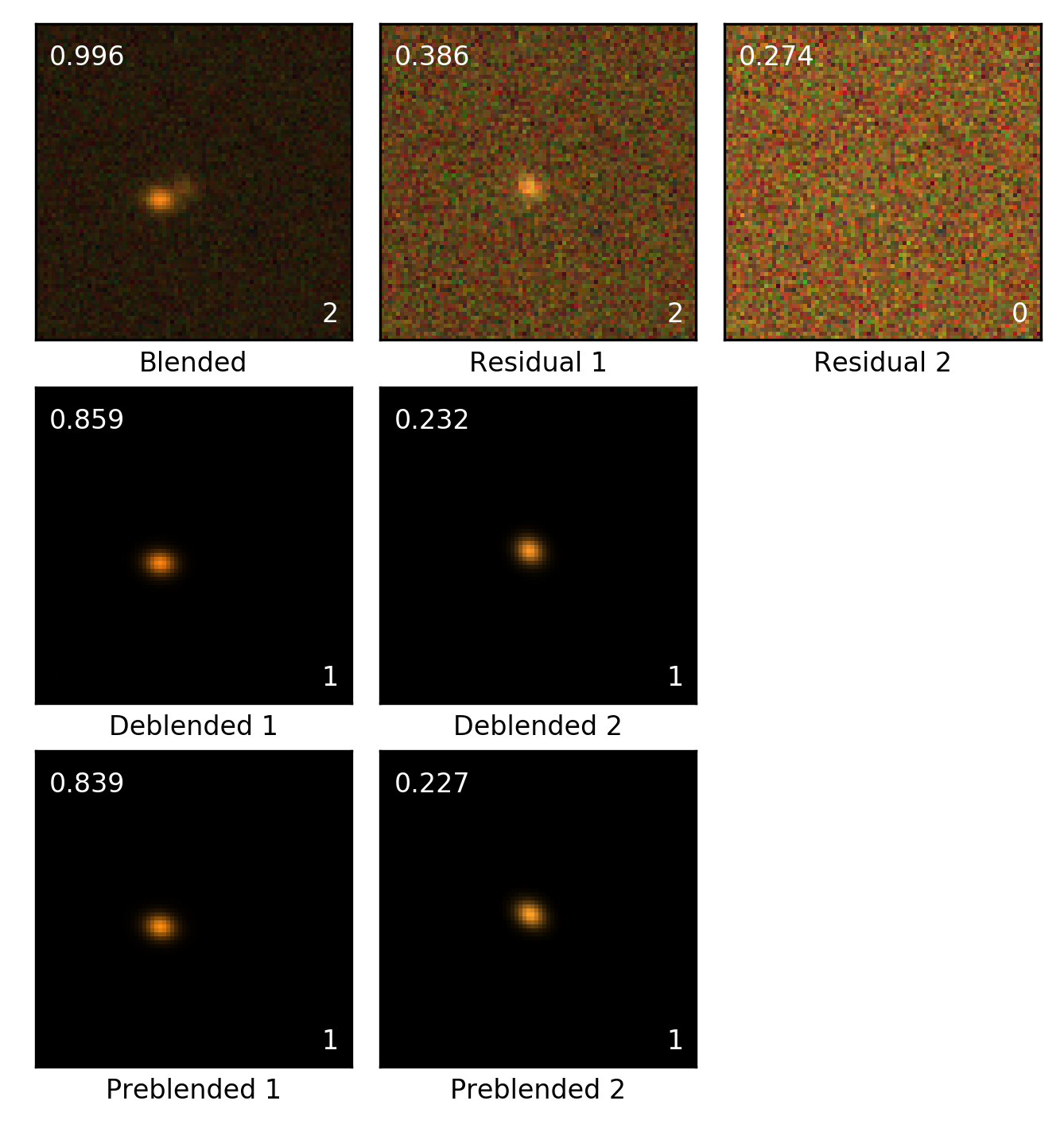}
    \includegraphics[width=\linewidth]{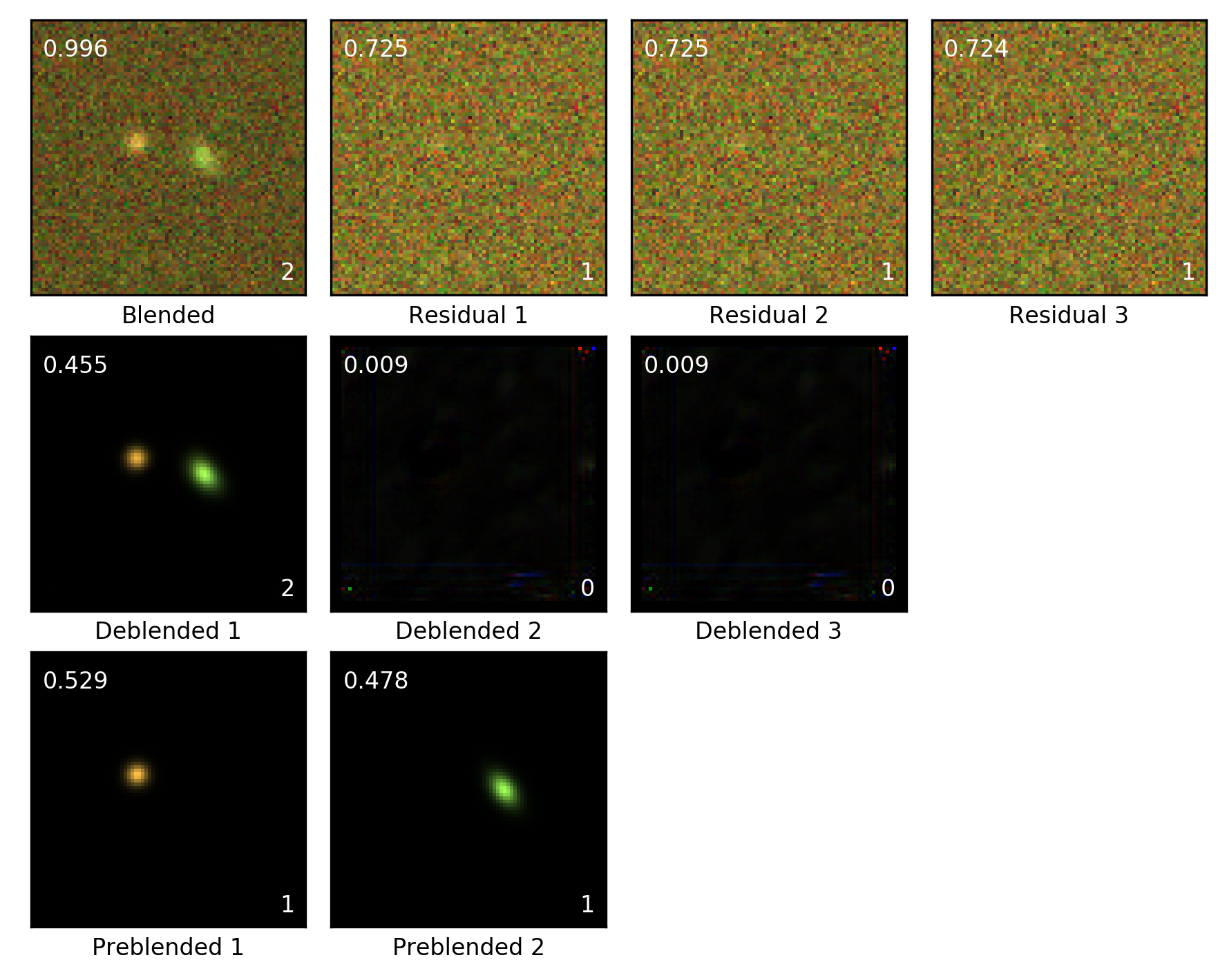}
    \caption{As FIG. \ref{fig:iterative_blend2}, but for medium and low quality samples.}
    \label{fig:iterative_blend2_medium_and_low}
\end{figure}

\begin{figure}[htbp]
    \centering
    \includegraphics[width=\linewidth]{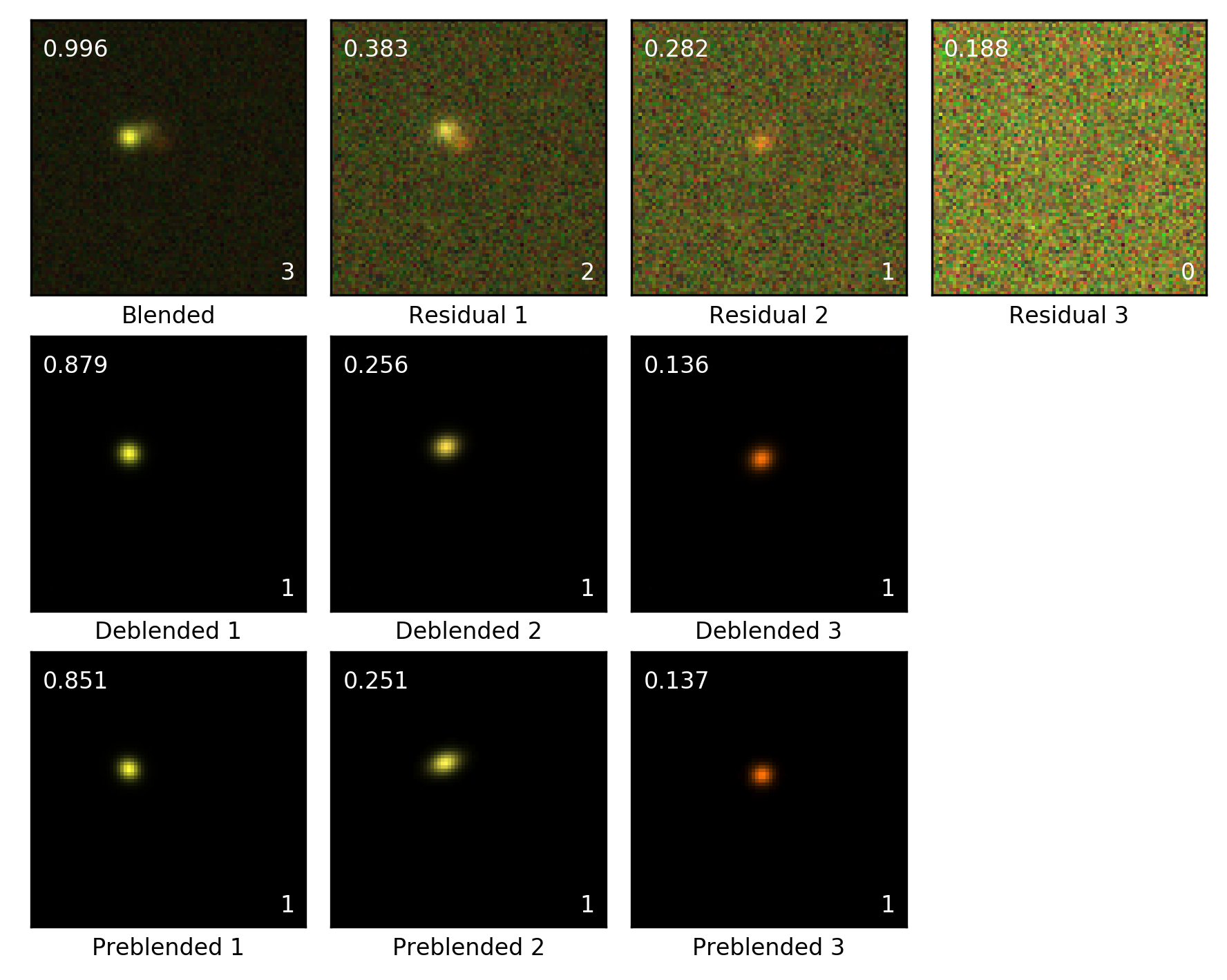}
    \includegraphics[width=\linewidth]{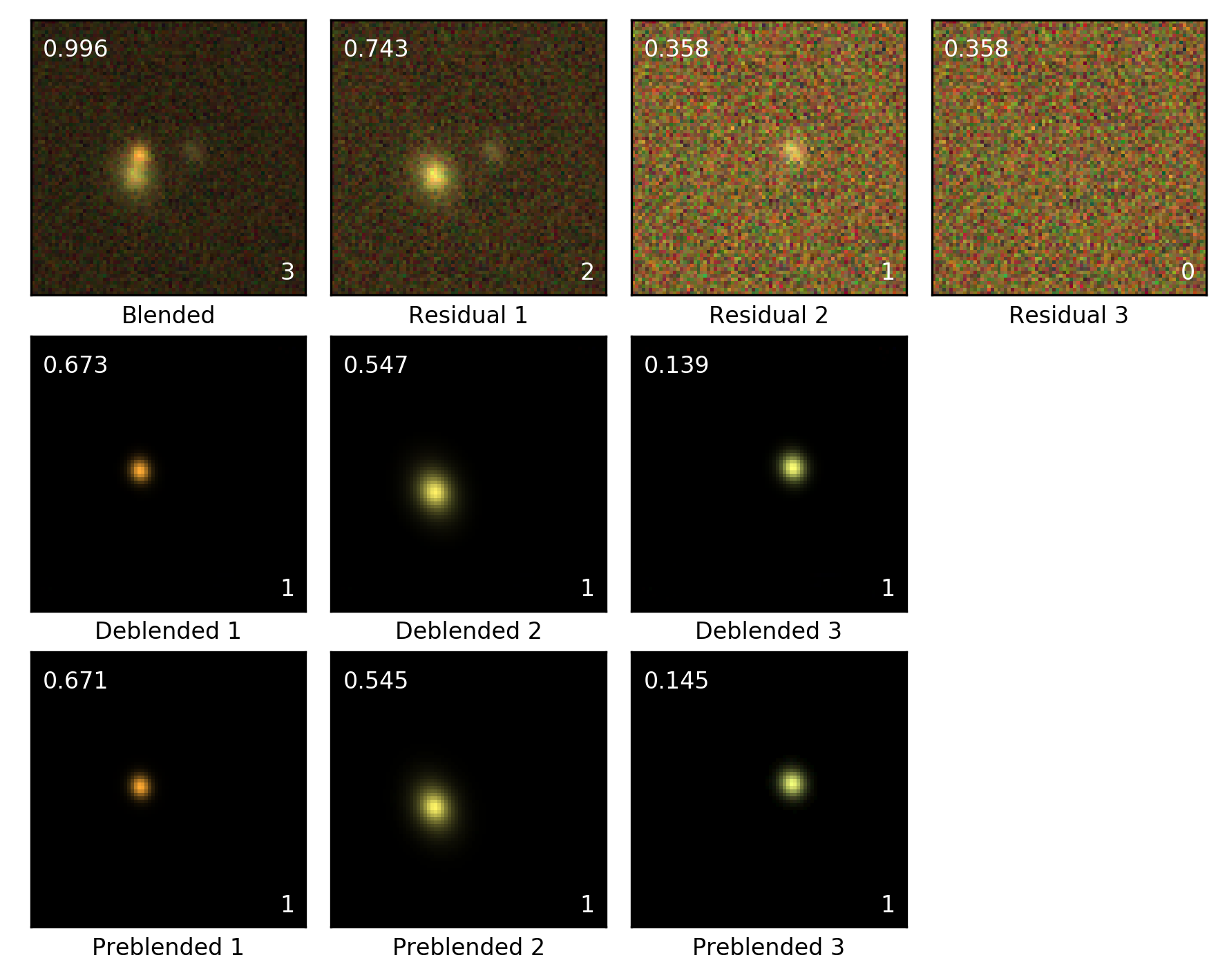}
    \caption{As FIG. \ref{fig:iterative_blend2}, but for $3$-galaxy blends.}
    \label{fig:iterative_blend3}
\end{figure}

\begin{figure}[htbp]
    \centering
    \includegraphics[width=\linewidth]{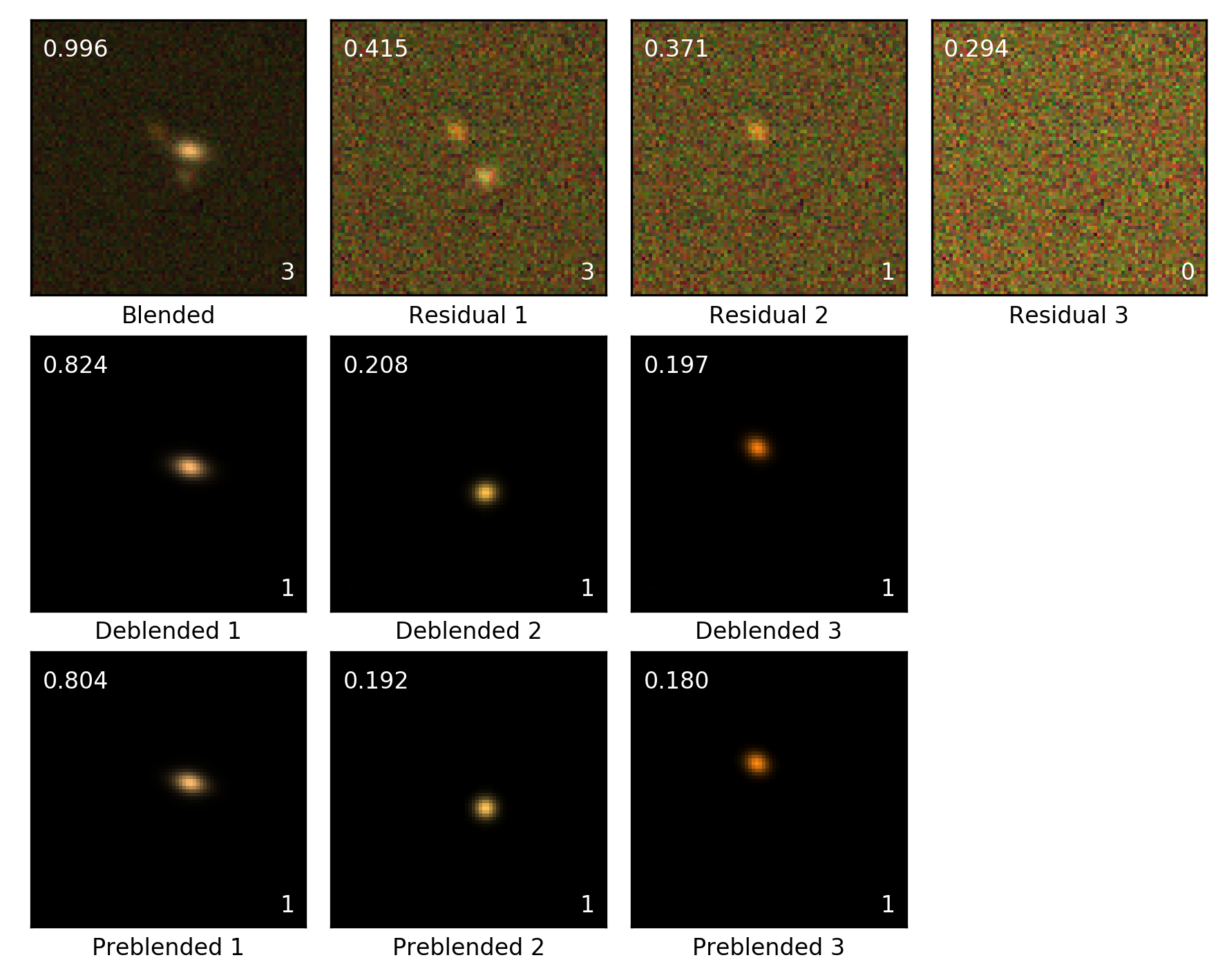}
    \includegraphics[width=\linewidth]{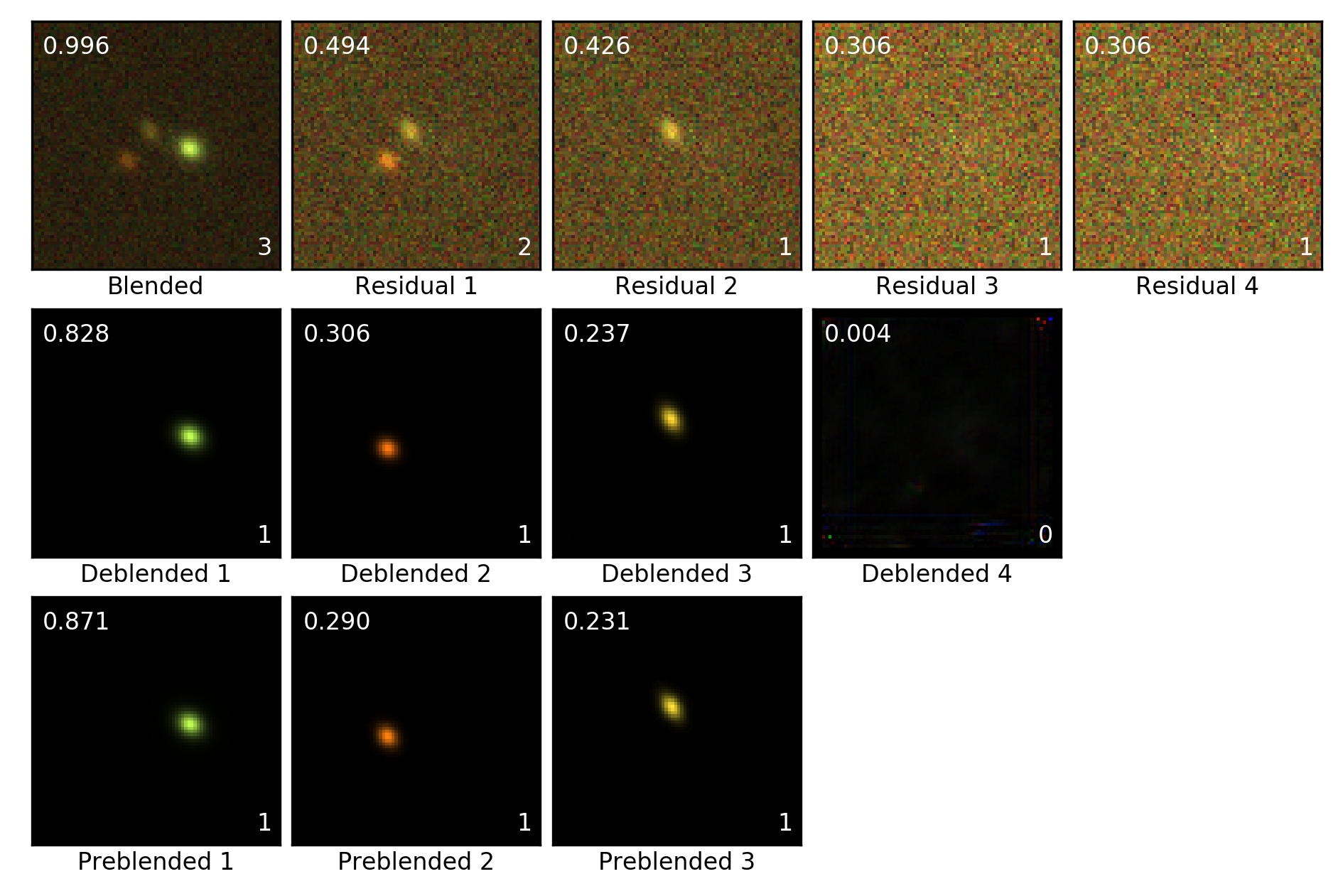}
    \caption{Figure showing medium and low quality samples for $3-$galaxy blends.}
    \label{fig:iterative_blend3_medium_and_low}
\end{figure}

\section{Results} \label{sec:Result}
\subsection{Experimental settings}
\subsubsection{Implementation details}
The RDN used in our framework contains $D=16$ RDBs with $C=8$ convolutional layers in each block. During the first phase of the training process, the learning rate for the RDN is $10^{-4}$, the batch size is $128$, and it is trained for $150$ epochs with Adam optimizer. For the classifier, the initial learning rate is $0.1$, the batch size is $200$, and it is trained for $200$ epochs. In the second phase, the deblender and the classifier are updated jointly, where the learning rate decays to $10^{-5}$ for the former and $10^{-6}$ for the latter. After some experimentation, we chose the trade-off coefficient $\lambda=10^{-4}$. The batch size is $8$ due to GPU limit and the framework is trained for $10$ epochs in this phase.

\subsubsection{Evaluation} \label{sec:evaluation}
We trained the framework on $2$-galaxy blended images and tested the trained model on 
blended images with $2$ to $7$ galaxies. Testing set for each class contains $1000$ noisy blended galaxy images.

We start by showing some example results from $2$-galaxy and $3$-galaxy blended images in FIG.s \ref{fig:iterative_blend2} and \ref{fig:iterative_blend3}. These results show the typical output from a high-quality deblend. The network is using both the morphological as well as the color information to isolate images of individual galaxies which is helpful in recovering images even in difficult cases.

\begin{table*}
\centering
\begin{tabular}{c|cccccc}
\hline \hline
 & 2-galaxy  & 3-galaxy & 4-galaxy & 5-galaxy & 6-galaxy & 7-galaxy\\
\hline 
High-quality & 77.3\%  & 50.4\%  & 40.6\%  & 27.2\%  & 14.7\%  & 10.5\%  \\
Medium-quality & 15.3\%  & 36.9\%  & 46.4\%  & 52.8\%  & 62.5\%  & 63.6\%  \\
Low-quality & 7.4\%  & 12.7\%  & 13.0\%  & 20.0\%  & 22.8\%  & 25.9\%  \\
\hline \hline
\end{tabular}
\caption{\label{tab:iterative_all} Table showing results of the iterative test on blended images with various number of galaxies.}
\end{table*}

In Table~\ref{tab:iterative_all} we present the fraction of deblends sorted by the quality category (see Section \ref{fwdesc}) for 
all the testing sets with various numbers of galaxies. Note that this only indicates how well the scheme thinks it is doing, rather than how well it is actually doing. With this caveat, the process shows relatively good results with a large fraction of nominally high-quality deblends. 

Somewhat surprisingly, the method continues to work even when the number of galaxies in the blend goes well beyond two as in training. Table~\ref{tab:iterative_all} shows a steady decrease in the number of high-quality deblends rather than a catastrophic drop in performance. A high-quality deblend for $7$-galaxy sample is presented in FIG. \ref{fig:iterative_blend7} in Appendix \ref{more_itr}. It seems that the network can deal with additional flux around the objects, regardless of whether it is coming from a single object or multiple objects.

\begin{table*}
\centering
\begin{tabular}{c|cc|cc|cc|cc}
  \hline \hline
  & \multicolumn{4}{c|}{2-galaxy problem} & \multicolumn{4}{c}{3-galaxy problem}  \\
  \hline
  & \multicolumn{2}{c|}{PSRN} & \multicolumn{2}{c|}{SSIM} & \multicolumn{2}{c|}{PSRN} & \multicolumn{2}{c}{SSIM}  \\
  \hline
  & Mean  & Median & Mean  & Median & Mean  & Median & Mean  & Median \\
\hline 
Deblended 1 & 56.51  & 57.93   & 0.9973  & 0.9997 & 54.66  & 57.19  & 0.9954  & 0.9997  \\
Deblended 2 & 58.18  & 59.02   & 0.9967  & 0.9997 & 55.96  & 58.77  & 0.9929  & 0.9996  \\
Deblended 3 &        &         &         &        & 57.71  & 59.29  & 0.9956  & 0.9995  \\
  \hline \hline
\end{tabular}

\caption{\label{tab:bigtable} Table showing PSNR(dB) and SSIM for $2$-galaxy and $3$-galaxy deblending problems.}

\end{table*}

To quantify the quality of the deblends, we start with some standard image analysis metrics. We applied the peak signal-to-noise ratio (PSNR) and the structural similarity index (SSIM) as metrics to evaluate the quality of deblended images from the deblender when compared with the ground truth.

PSNR represents a peak error in the measurement of reconstruction quality in image compression. It computes the logarithm of the ratio between the maximum pixel value (MAX) of the ground truth and the mean squared error (MSE) between the test image and the ground truth in decibels. In our experiment, the ground truth consists of noiseless images with a single galaxy and the test images are the corresponding deblended images from the RDN. PSNR is formulated as Eq.~\ref{eq:PSNR}:

\begin{equation}
    {\rm PSNR(dB)} = 20\cdot \log_{10}({\rm MAX})-10\cdot \log_{10}({\rm MSE}) \label{eq:PSNR}
\end{equation}

SSIM~\cite{wang2004image}, known as the structural similarity index, is commonly used to evaluate the similarity between two images using the means $\mu_x$ and $\mu_y$, the variances $\sigma_x$ and $\sigma_y$ and the covariance $\sigma_{xy}$. In Eq.~\ref{eq:SSIM}, $c_1=(k_1L)^2$ and $c_2=(k_2L)^2$ are two small constants to avoid the instability with a weak denominator. We use $k_1=0.01$, $k_2=0.03$ by default and $L$ is the dynamic range of pixel values.

\begin{equation}
    {\rm SSIM} = \frac{(2\mu_x\mu_y+c_1)(2\sigma_{xy}+c_2)}{(\mu_x^2+\mu_y^2+c_1)(\sigma_x^2+\sigma_y^2+c_2)} \label{eq:SSIM}
\end{equation}

Results are summarized in Table \ref{tab:bigtable} and Table \ref{tab:bigtable_extension} in Appendix \ref{more_itr}. We do observe a few trends here. First, the PSNR of the second deblended galaxy is higher than that of the first. This somewhat counter-intuitive result comes from the normalization of PSNR, since the second deblend is fainter than the first one. In other words, the PSNR is not telling us that the 2nd deblended galaxy is better deblended, only that its quality is less degraded than expected, given how much fainter it is. This interpretation is confirmed by the SSIM values, which are higher for the first deblend, but only marginally. In both cases we see that the $3$-galaxy problem is more difficult than the $2$-galaxy problem. We also see that the median values are systematically above the mean values, telling us that the mean is pulled down by a small number of catastrophic outliers. To study these results in language that is more relevant to the astronomy community, we turn to the recovery of the fluxes and moments, described in the next section.

\subsection{Flux and image moment recovery using RDN}
\label{sec:flux-image-moment}

\begin{table*}
    \centering
    \begin{tabular}{c c | c c | c c c }
     \hline
     \hline
     sample & quality     &  $\alpha_f$ & $\sigma^2_f$ & $|\alpha_e|$ & ${\rm Arg}(\alpha_e) $ &  $\sigma^2_e$ \\
     \hline
    x   & high      & $1.008\pm0.009$ & $1.1$ & $0.52 \pm 0.04$ & $-0.01 \pm 0.04$ & $0.034$ \\
    x   & medium    & $1.029\pm0.014$ & $6.0$ & $0.56 \pm 0.08$ & $-0.08 \pm 0.12$ & $0.046$ \\
    x   & low       & $1.014\pm0.014$ &$2.2$ & $0.65 \pm 0.12$ & $-0.15 \pm 0.14$ & $0.049$ \\
    \hline
    y   & high      & $1.008\pm0.010$ &$1.2$ & $0.38 \pm 0.04$ & $0.02 \pm 0.05$ & $0.024$ \\
    y   & medium    & $0.959\pm0.032$ &$2.4$ & $0.50 \pm 0.07$ & $0.19 \pm 0.10$ & $0.026$ \\
    y   & low       & $0.52\pm0.14$ &$6.7$ &  $0.18 \pm 0.13$ & $1.20 \pm 1.09$ & $0.084$\\
    \hline
    \hline
    \end{tabular}
    \caption{Table showing the recovered response factors for flux and galaxy ellipticities split by the deblend type (\texttt{x} or \texttt{y}) and quality flag. Errors on responses were derived by subsampling using bootstrap technique with replacement.  }
    \label{tab:props}
\end{table*}

\begin{figure*}
    \includegraphics[width=0.8\linewidth]{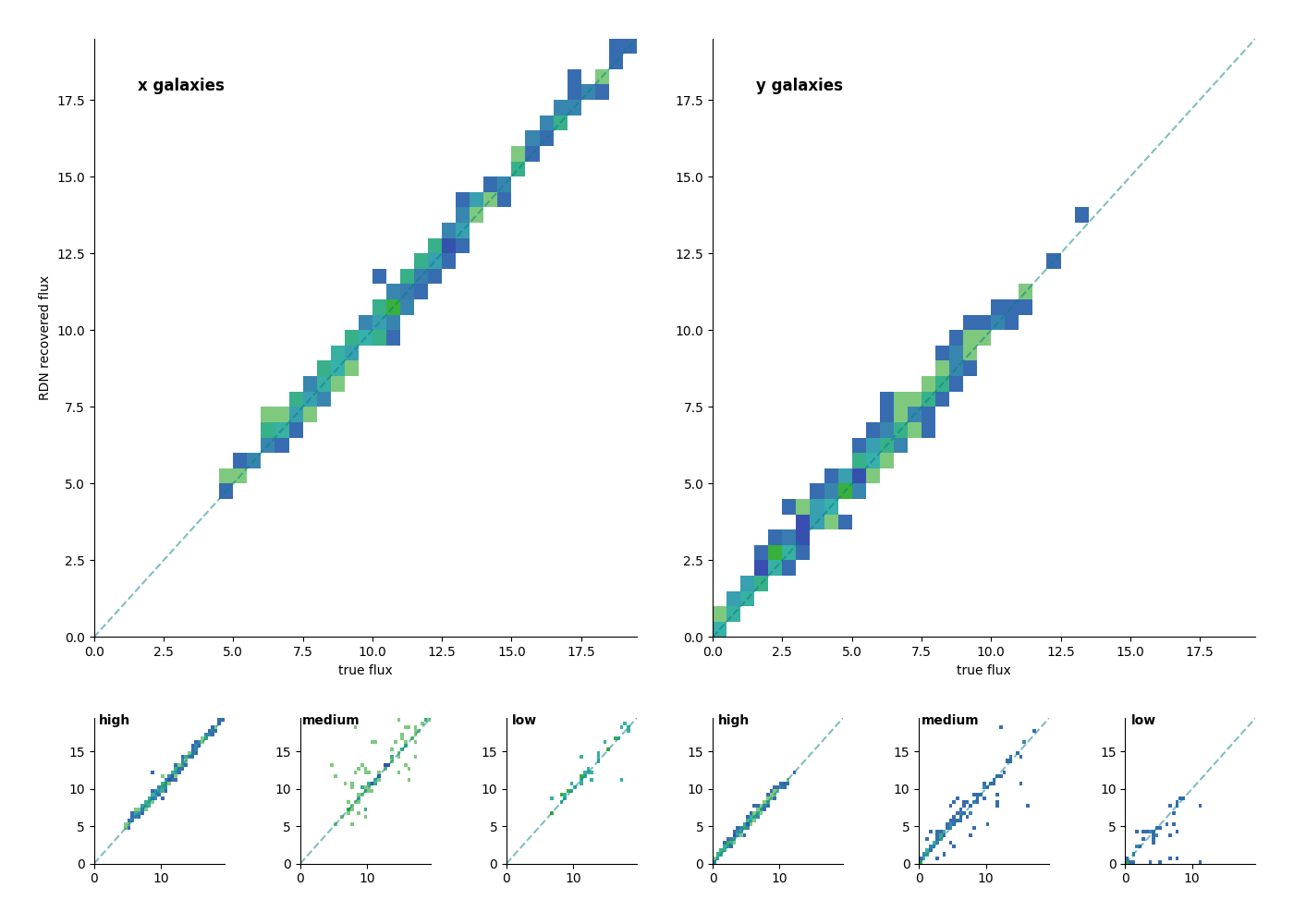}
     \captionsetup{justification=justified}
    \caption{2-d histograms showing the comparison of the fluxes recovered from the RDN and the true fluxes for the brighter \texttt{x} galaxies and the fainter \texttt{y} galaxies. The teal line is $y=x$ expectation. The smaller plots show the same relation separated in quality bins of the image data.}
    \label{fl_xcomp_SE}
\end{figure*}

An important measure of the effectiveness of any algorithm such as ours is the accurate recovery of physical parameters from the input image. To that end, we have opted to compare the fluxes and the second order image moments (which represent the equivalent ellipse of the image and by extension, the shape and orientation) retrieved by the RDN against that of the truth images. For ease of reference, the brighter galaxies in each field are referred to as \texttt{x} galaxies and the fainter ones as \texttt{y} galaxies.
 
FIG. \ref{fl_xcomp_SE} shows the RDN recovered fluxes (y-axis) plotted against the true fluxes (X-axis) for the \texttt{x} and \texttt{y} galaxies for the entire test set. The larger panels show the distribution for the entire test set while the smaller panels at the bottom show the distribution separated by quality as detailed in Section \ref{fig:framework}. The dotted teal line is the $y=x$ expectation.

As is evident from FIG. \ref{fl_xcomp_SE}, the RDN is very efficient in recovering the fluxes for the brighter galaxies in our fields, more or less uniformly so for the high, medium and low quality images. 
 As expected, the lower quality assignments are associated with a higher scatter between real and measured fluxes, although for the brighter galaxy, there is no measurable difference between the medium and low quality marks (given the size of our test catalog). For the fainter galaxies, however, the low quality deblends are essentially uncorrelated with the true values. We find that except for secondary objects in the low-quality bin, our RDN based estimator is essentially unbiased in recovering flux values.

To quantify this we model the recovered flux $f_r$ as 
\begin{equation}
    f_r = \alpha_f f_t + n_f
\end{equation}
where $\alpha_f$ is a proportionality constant and $n_f$ is flux noise which has zero mean and variance $\sigma^2_f$. The maximum likelihood estimate for $\alpha_f$ and $\sigma^2_f$ are given by
\begin{eqnarray}
    \alpha_f &=& \frac{\sum f_r f_t}{\sum f_t^2} \\
    \sigma^2_n &=& \frac{1}{N} \sum (\alpha_f f_r - f_t)^2
\end{eqnarray}
In Table \ref{tab:props} we show the values of the recovered parameters. We see that in the case of flux, $\alpha_f$ is consistent with unity for both \texttt{x} and \texttt{y} galaxies indicating unbiased flux measurements,  except for the low quality \texttt{y} regime, where deblending suffers from bad systematic errors.
We also observe from the table that the variance/noise $\sigma_f^2$ increases as the quality of the blends decreases, except in the case of medium quality \texttt{x} galaxies, which is likely because of some outliers (as evidenced by the scatter for the respective quality type in FIG. \ref{fl_xcomp_SE}). 


Similar to the fluxes, the shapes and orientations of the galaxies, represented by their second order image moments, are crucial characteristics that we aimed to recover using the RDN, in particular with the weak lensing application in mind. We define second moments as

\begin{eqnarray}
    \mu_{ij} = \frac{\sum_{xy} x^i y^j I(x,y)}{\sum_{x,y} I(x,y)} \quad  &\mbox{for\ }& i+j=1 \nonumber\\
    \mu_{ij} = \frac{\sum_{xy} (x-\mu_{10})^i (y-\mu_{01})^j I(x,y)}{\sum_{xy} I(x,y)} \quad &\mbox{for\ }& i+j=2 \nonumber
\end{eqnarray}
where the sum goes over image pixels,  $I$ is the pixel intensity and $\bar{x}$ and $\bar{y}$ are the mean values. The first moments measure the central position of the stars, while the second moments measures the shape. Commonly used shape parameters are ellipticities, given by
\begin{eqnarray}
e_1 &=& \frac{\mu_{20} -\mu_{02}}{\mu_{20}+\mu_{02}} \\
e_2 &=& \frac{2 \mu_{11}}{\mu_{20}+\mu_{02}}
\end{eqnarray}
The ellipticities so defined are dimensionless and are invariant both under scaling in flux and size. We are calculating these quantities for true and neural network recovered images, which are noiseless (i.e. flux is zero outside object's extent) and so there is no need for more advanced techniques such as adaptive moments.

\begin{figure*}
\includegraphics[width=\linewidth]{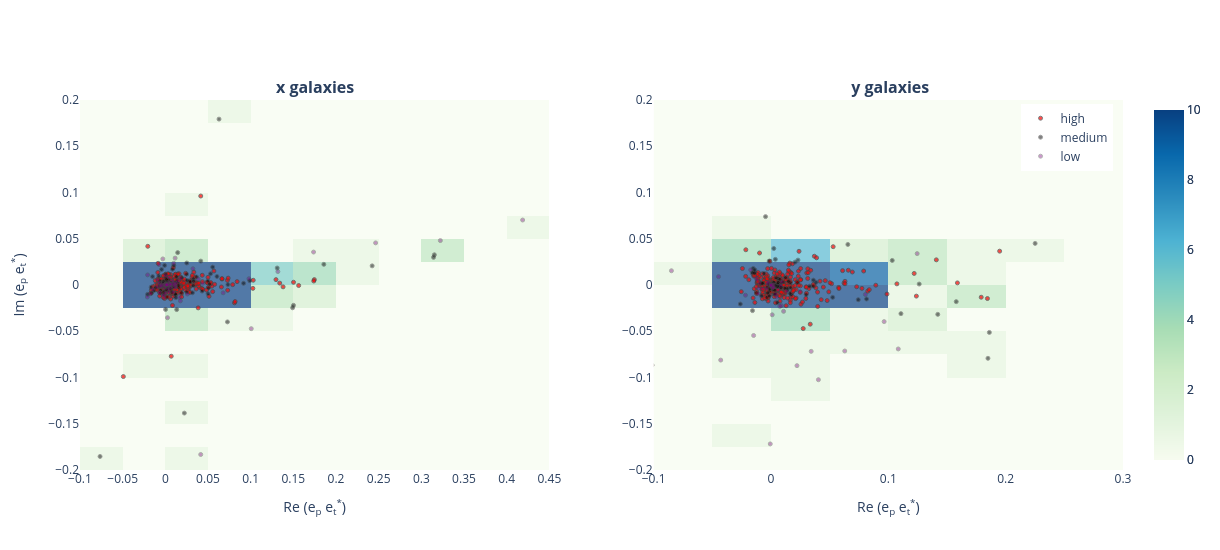} 
\captionsetup{justification=justified}
\caption{ 2-d histograms showing the quantity $e_p e_t^*$ plotted on the complex plan. For a noiseless recovery, the points would scatter on the positive $x$-axis line weighted by the square of true ellipticity. The left plot is for the \texttt{x} galaxies and the right plot is for the \texttt{y} galaxies. Different quality flags are represented by different colored points as per the legend. The scatter around this line represents the level of the noise in moment recovery. See table \ref{tab:props} for a quantification of the deblending response.}
\label{debtr_e_2dhist}
\end{figure*}

In order to plot a two dimensional quantity such as ellipticity against the truth, we plot the value of $e_p e_t^*$ on the complex plane in the Figure \ref{debtr_e_2dhist}. This in turn "rotates" ellipticity by the inverse of its true argument, and in the ideal case, the results should scatter around the positive real axis. Additionally, dividing by $e_t$ blows up the value for galaxies with small ellipticities making the plot unwieldy. We see that with the exception of a few catastrophic outliers, the galaxies indeed do scatter around the real axis. The y galaxies are noisier, but neither is biased.

In order to quantify the shape recovery we fit a similar model as in the case of flux:
\begin{equation}
    f_e = \alpha_e e_t + n_e
\end{equation}
Note that since $e_t$ is now a complex quantity, $\alpha$ is also a complex quantity. Results of this exercise are shown in the Table \ref{tab:props}. For an ideal recovery, $\alpha_e=1$ and real. Since the noise is more likely to scatter object into a rounder shape, $\alpha_e<1$. We find $\alpha_e\sim 0.55$ for the \texttt{x} galaxies seemingly independent of the quality flag. Average $\alpha_e$ for \texttt{y} galaxies is even lower. The argument of $\alpha_e$ is is consistent with zero except for the low quality \texttt{y} galaxies. This indicates that the recovery is biased in amplitude, as is expected for quantities that are non-linear in image fluxes, but not in orientation. Only in the case of low quality \texttt{y} galaxies, the results indicate  that the output shapes are uncorrelated with the input shapes.  We also find that the value of noise, $\sigma_e^2$ increases with the quality flag as expected.

In an earlier version of this work we have found cases where the recovered galaxies would be drastically different than input cases: the neural network would sometimes turn round objects into highly elliptical ones or vice-versa. We have manually looked at several of these cases in the Appendix \ref{appc} without identifying a clear regularity.

\subsection{Comparison with SExtractor}
\label{sec:comp-with-sextr}

\begin{figure*}
    \captionsetup{justification=justified}
    \includegraphics[width=0.8\linewidth]{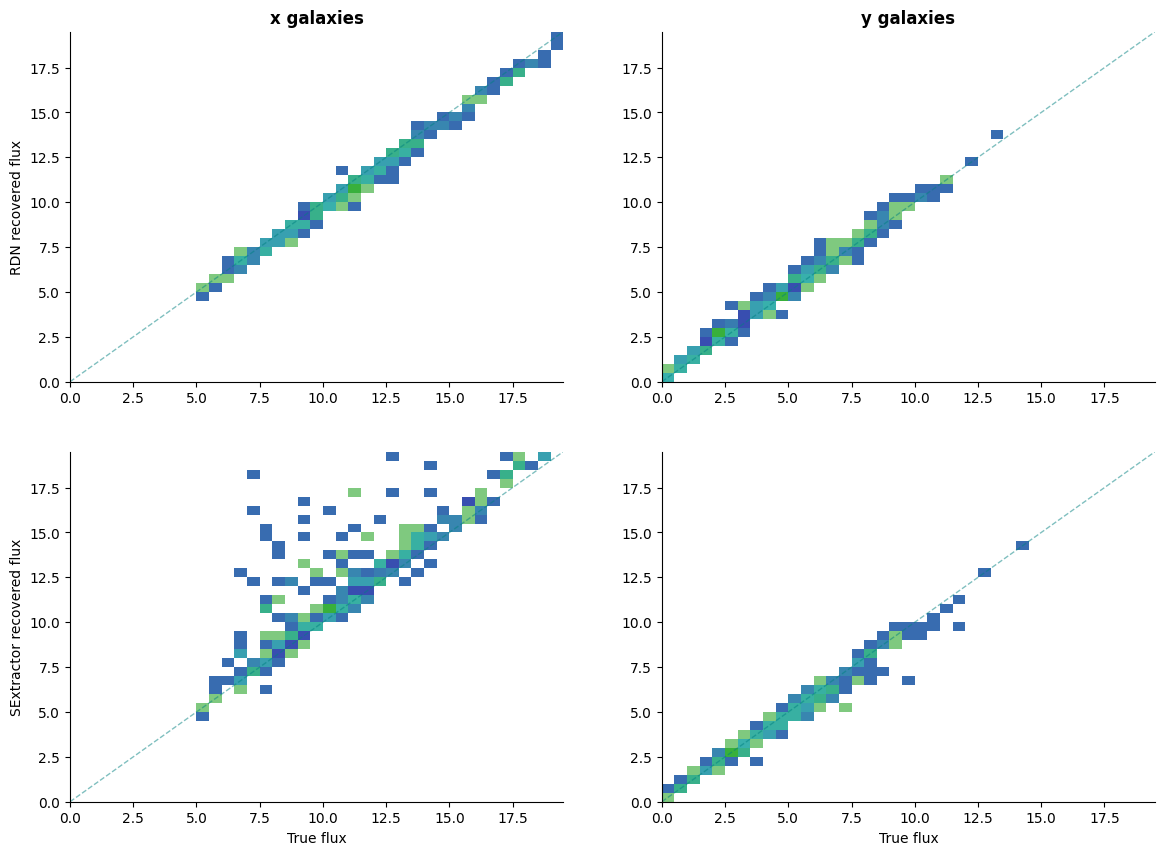}
    \caption{2-d histograms showing the flux recovery from RDN and SExtractor as compared to the true fluxes. The top left and right panels show the true fluxes on the X-axis and the RDN  fluxes on the y-axis respectively for \texttt{x} \& \texttt{y} galaxies. The teal line is the $y=x$ expectation. The bottom left and right panels are the same representations for SExtractor.}
    \label{rdnsexcomp_flux}
\end{figure*}

In this section, we compare our deblending strategy with what is widely considered as the industry standard, Source Extractor (SExtractor, \cite{B&A1996}), which has been the baseline detection, deblending and image extraction software in astronomy for over two decades. It returns a set of user specified parameters from a more extensive default parameter file, by following the specific configurations defined by the user. In this work, we have used the Python formulation of SExtractor, \textit{sep} \cite{sep2015} with the settings employed for DES\footnote{\url{https://github.com/esheldon/sxdes}}.

We note that this is a fundamentally unfair comparison for several reasons. First, the RDN has been trained for this particular set of galaxies and thus internally employs a correctly tuned prior for the distribution of morphologies, fluxes and ellipticities, while SExtractor employs a general algorithm and is thus intrinsically more robust. While the RDN's approach to deblending is peeling off the individual objects, SExtractor detects objects by creating segmentation maps. Moreover, following DES configuration, SExtractor is run on the co-added image (R+G+B) and is thus missing the color information. It is also important to note that although SExtractor outputs a number of photometric features, it is primarily used to segment image while a detailed analysis is left to other codes. Nevertheless, it is an appropriate sanity check to measure what kind of improvements can be brought about by employing more sophisticated methods.

FIG. \ref{rdnsexcomp_flux} shows the comparison of the flux recovery by the RDN and SExtractor for both the \texttt{x} and \texttt{y} galaxies. The top left and right panels feature the RDN recovered fluxes on the y-axis and the true fluxes on the X-axis for the \texttt{x} and the \texttt{y }galaxies respectively. The bottom left and right panels are the same representations for the SExtractor fluxes. The teal line is a $y=x$ fit. As is evident from the plots, the RDN does a better job of recovering the object fluxes for both the \texttt{x} and \texttt{y} galaxies.  SExtractor has a tendency to put a disproportionate amount of flux into the first detected objects, and hence the \texttt{x} images are biased high, while the \texttt{y} images are biased low.

\subsection{Failure modes}

In this section we investigate the most common failure modes in the two approaches by cherry-picking cases in which one method performs correctly while the other fails.

\begin{figure*}
    \includegraphics[width=0.65\linewidth]{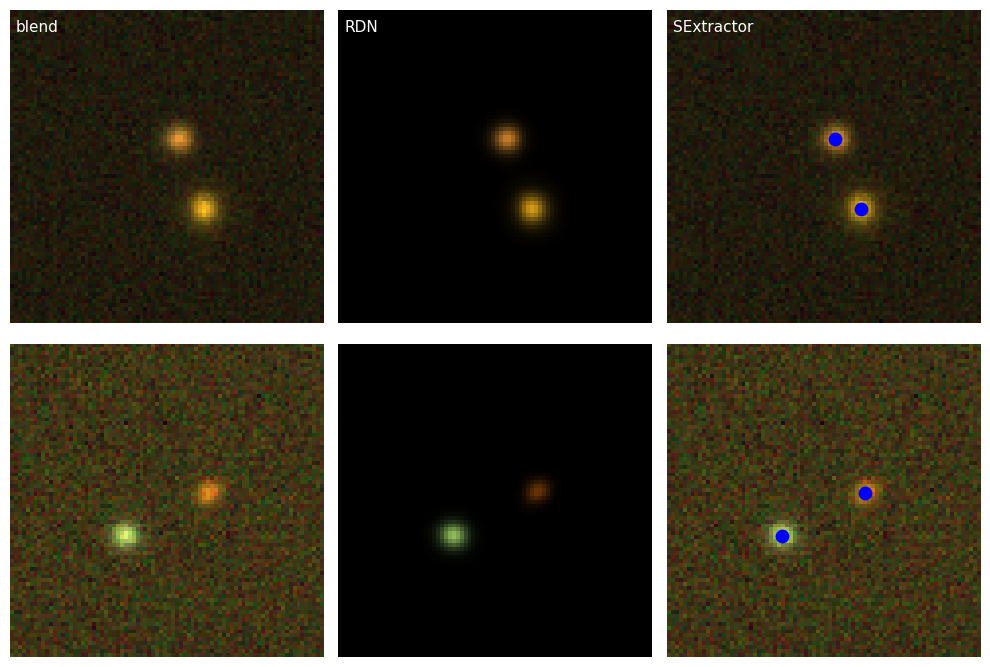}
    
     \captionsetup{justification=justified}
    \caption{Figure showing instances where the RDN fails, while SExtractor succeeds in detecting and deblending the objects in the field. On the left is the original blend, in the centre is the output of the RDN and on the right is the blend with the SExtractor detections superimposed in blue.}
    \label{rdnfail_1}
\end{figure*}

We manually looked at all the cases where the RDN detects just a single galaxy. In a majority of those, the secondary object was very faint and overlapped almost perfectly with the primary, leading to the deblending being suspended because the \classifier\ determined that there are no further objects. However, for $12$ out of the $1000$ objects in the test set, for the deblended \texttt{x} galaxies, the result clearly shows two objects, examples of which are shown in FIG. \ref{rdnfail_1}. These problems are trivial for the SExtractor approach. A visual inspection of all these images was conducted and it was observed that, both the objects in $7$ of these fields are very similar in shape and, in $5$ fields, they had similar brightnesses. It stands to reason that the RDN has some difficulty in deblending objects with similar structural parameters, most likely because it cannot decide which the ``brighter'' object is, especially in the presence of noise. In fact, when presented with such examples in training, it simply pays a large loss penalty whenever it starts with the ``wrong'' object, without a clear rule about how to pick one of the two objects if the noise hides the identity of the bright ones. We will discuss this further in Section \ref{sec:Conclusion}.

\begin{figure*}   
    \includegraphics[width=0.8\textwidth]{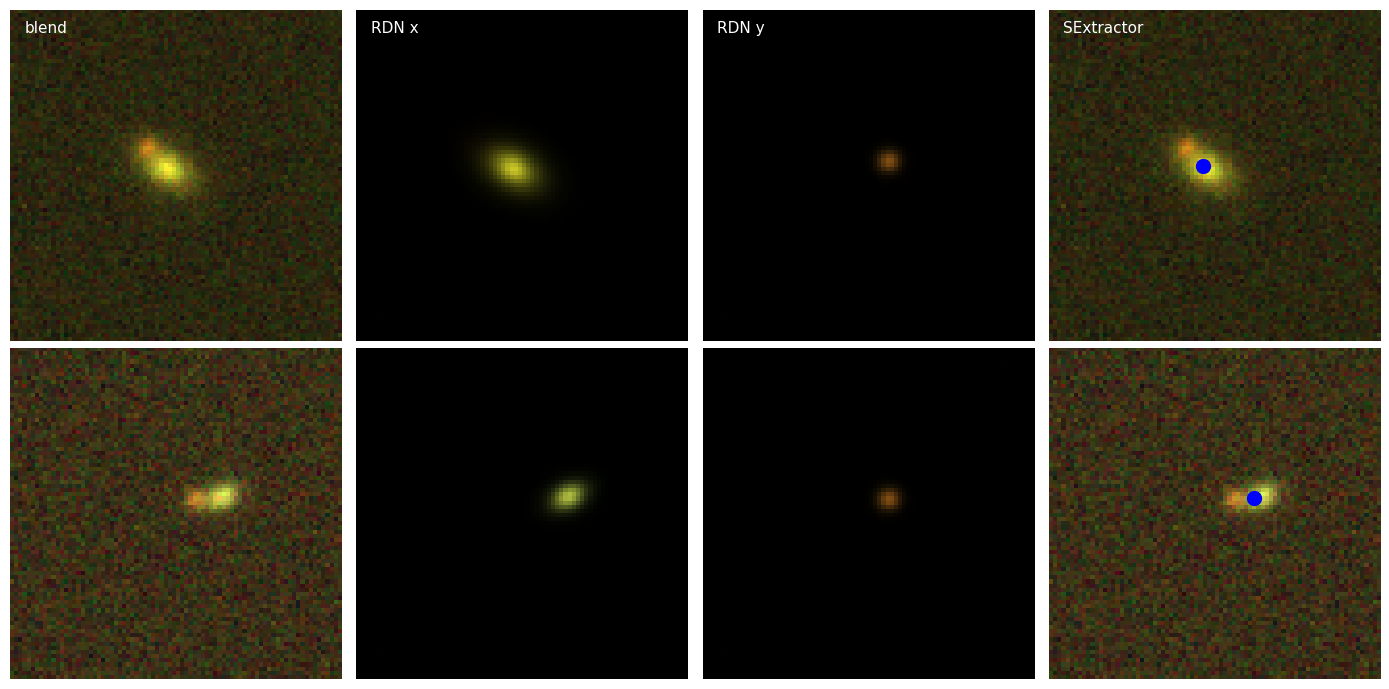}
    
     \captionsetup{justification=justified}
    \caption{Figure showing instances where SExtractor fails to detect and deblend, while the RDN succeeds. On the far left is the original blend, the RDN result is represented in the two central panels for the \texttt{x} and \texttt{y} galaxies, and the blend with the SExtractor detection superimposed in blue is on the far right panel.}
    \label{sexfail_1}
\end{figure*}

When SExtractor is run on the test set, it returns photometric parameters for $979$ \texttt{x} galaxies and  $729$ \texttt{y} galaxies (the \texttt{x} counterparts of all \texttt{y} galaxies are present in the \texttt{x} set). As seen in FIG. \ref{rdnfail_1}, this includes $9$ instances where the RDN fails to deblend the different galaxies in the field. However, overall, there are more instances where the RDN successfully deblends the \texttt{x} and \texttt{y} galaxies, while SExtractor either doesn't detect either galaxy, or only detects the  \texttt{x} galaxy, but not the the \texttt{y} galaxy, a few examples of which are illustrated by FIG. \ref{sexfail_1}. We note that these are traditional blend merges, where the deblender merges two distinct objects into a single one. Since human eye is pretty good at detecting these, we could possibly improve upon this by fine-tuning the SExtractor settings, potentially at the cost of artificially shredding objects with substructure. These are also examples where color information is most helpful.

\section{Summary and Conclusions} \label{sec:Conclusion}

In this work we present a new approach to deblending using a Residual Dense Network, which was trained on blended galaxy images with a realistic PSF and noise levels, and which has performed decently in deblending and recovering object fluxes and shapes. Compared to previous works, our set up does not assume that the object to be deblended is located at the center of the image. We also do not need to assume in advance, the number of objects to be deblended. By using two neural networks, one to perform deblending and one to determine the number of remaining objects in the field, we can classify the quality of deblends. We have shown that deblends that are designed to be higher quality have indeed, less noisy fluxes and shapes.

In the most current deblending networks found in literature, the network is trained to deblend the central galaxy. Given that centering in the presence of noise presents its own problems, we have trained our network to deblend the brightest galaxy remaining in the image. This works very well, where the choice of the brightest is clear, but leads to a particular failure mode where the network cannot decide where to start, when two objects are approximately equally bright (see FIG. \ref{rdnfail_1}). This might be a generic feature of any kind of deblender that proceeds by peeling off one galaxy at a time. For example, even if we choose the galaxy closer to the center, there would likely be an equivalent failure mode when two galaxies are equally close to the center. The same thing could happen for other initial assumptions that attempt to distinguish between the galaxies. There are several approaches to solve this. One possibility would be to use a symmetrized loss function that would train the network to deblend either galaxy. Another possibility would be to have a more sophisticated network that deblends multiple galaxies at once, perhaps again with an appropriately symmetrized loss function.

Despite the RDN deblender being trained on two-galaxy blends, it performs surprisingly stably all the way to the maximum 7-galaxy blends, where it still correctly deblends 7 galaxies in 10\% of cases. 

We have found that our method outperforms the industry standard SExtractor on flux recovery, subject to the strong caveats mentioned in Section \ref{sec:comp-with-sextr}. However, we found that the RDN recovered shapes are significantly worse and occasionally catastrophically wrong. The very good results could partially be the result of the limited morphological variety of the presented images; however, since we do not attempt to deconvolve the image, this might not be a major issue. Our work reiterates concerns with using a deeply non-linear deblender based on NN for work requiring precision shapes, such as weak lensing. At the same time, such work could lead to potentially interesting new work on galaxy-star separation and morphological classification of galaxies.

Similar to the majority of other neural network approaches, our deblender cannot deal optimally with a PSF that is variable and is different for the different input channels (for e.g. for different bands), nor can it deal with a pixel mask. This is a significant deficiency that is usually swept under the rug under the assumption that the network can be retrained for a set of different PSFs. However, in practice, we have a different PSF for different bands and with the typical astronomical observations in five or six bands, the number of possible PSF combinations becomes unmanageable. The PSF shape in each band thus needs to be a part of the input and network training. We are leaving these important aspects for future work.

One of the main advantages of most ML approaches to deblending is that they also automatically denoise and deconvolve the image and can in principle fill in the missing information in case of pixel masks.  For a sufficiently sophisticated neural network, this would trivialize many typical operations such as flux and shear estimation on the resulting truth images. However, it would also require us to have a more advanced way of propagation of both statistical and systematic uncertainties. 

This work has demonstrated the feasibility of using RDN in astronomical image analysis and has highlighted some of the unresolved issues with the current machine learning approaches to the deblending problem in the case of realistic astronomical images.

\section*{Acknowledgements}
This material is based upon work supported by the U.S. Department of Energy, Office of Science, Office of Advanced Scientific Computing Research through the SciDAC grant ``Accelerating HEP Science: Inference and Machine Learning at Extreme Scales'' under DOE Contract Number DE-SC-0012704. We thank our collaborators for useful input.

\bibliography{references.bib}

\begin{thebibliography}{20}%
\makeatletter
\providecommand \@ifxundefined [1]{%
 \@ifx{#1\undefined}
}%
\providecommand \@ifnum [1]{%
 \ifnum #1\expandafter \@firstoftwo
 \else \expandafter \@secondoftwo
 \fi
}%
\providecommand \@ifx [1]{%
 \ifx #1\expandafter \@firstoftwo
 \else \expandafter \@secondoftwo
 \fi
}%
\providecommand \natexlab [1]{#1}%
\providecommand \enquote  [1]{``#1''}%
\providecommand \bibnamefont  [1]{#1}%
\providecommand \bibfnamefont [1]{#1}%
\providecommand \citenamefont [1]{#1}%
\providecommand \href@noop [0]{\@secondoftwo}%
\providecommand \href [0]{\begingroup \@sanitize@url \@href}%
\providecommand \@href[1]{\@@startlink{#1}\@@href}%
\providecommand \@@href[1]{\endgroup#1\@@endlink}%
\providecommand \@sanitize@url [0]{\catcode `\\12\catcode `\$12\catcode
  `\&12\catcode `\#12\catcode `\^12\catcode `\_12\catcode `\%12\relax}%
\providecommand \@@startlink[1]{}%
\providecommand \@@endlink[0]{}%
\providecommand \url  [0]{\begingroup\@sanitize@url \@url }%
\providecommand \@url [1]{\endgroup\@href {#1}{\urlprefix }}%
\providecommand \urlprefix  [0]{URL }%
\providecommand \Eprint [0]{\href }%
\providecommand \doibase [0]{https://doi.org/}%
\providecommand \selectlanguage [0]{\@gobble}%
\providecommand \bibinfo  [0]{\@secondoftwo}%
\providecommand \bibfield  [0]{\@secondoftwo}%
\providecommand \translation [1]{[#1]}%
\providecommand \BibitemOpen [0]{}%
\providecommand \bibitemStop [0]{}%
\providecommand \bibitemNoStop [0]{.\EOS\space}%
\providecommand \EOS [0]{\spacefactor3000\relax}%
\providecommand \BibitemShut  [1]{\csname bibitem#1\endcsname}%
\let\auto@bib@innerbib\@empty
\bibitem [{\citenamefont {{LSST Science Collaboration}}\ \emph
  {et~al.}(2009)\citenamefont {{LSST Science Collaboration}}, \citenamefont
  {{Abell}}, \citenamefont {{Allison}}, \citenamefont {{Anderson}},
  \citenamefont {{Andrew}}, \citenamefont {{Angel}}, \citenamefont {{Armus}},
  \citenamefont {{Arnett}}, \citenamefont {{Asztalos}}, \citenamefont
  {{Axelrod}},\ and\ \citenamefont {et~al.}}]{0912.0201}%
  \BibitemOpen
  \bibfield  {author} {\bibinfo {author} {\bibnamefont {{LSST Science
  Collaboration}}}, \bibinfo {author} {\bibfnamefont {P.~A.}\ \bibnamefont
  {{Abell}}}, \bibinfo {author} {\bibfnamefont {J.}~\bibnamefont {{Allison}}},
  \bibinfo {author} {\bibfnamefont {S.~F.}\ \bibnamefont {{Anderson}}},
  \bibinfo {author} {\bibfnamefont {J.~R.}\ \bibnamefont {{Andrew}}}, \bibinfo
  {author} {\bibfnamefont {J.~R.~P.}\ \bibnamefont {{Angel}}}, \bibinfo
  {author} {\bibfnamefont {L.}~\bibnamefont {{Armus}}}, \bibinfo {author}
  {\bibfnamefont {D.}~\bibnamefont {{Arnett}}}, \bibinfo {author}
  {\bibfnamefont {S.~J.}\ \bibnamefont {{Asztalos}}}, \bibinfo {author}
  {\bibfnamefont {T.~S.}\ \bibnamefont {{Axelrod}}},\ and\ \bibinfo {author}
  {\bibnamefont {et~al.}},\ }\bibfield  {title} {\bibinfo {title} {{LSST
  Science Book, Version 2.0}},\ }\href@noop {} {\bibfield  {journal} {\bibinfo
  {journal} {arXiv e-prints}\ ,\ \bibinfo {eid} {arXiv:0912.0201}} (\bibinfo
  {year} {2009})},\ \Eprint {https://arxiv.org/abs/0912.0201} {arXiv:0912.0201
  [astro-ph.IM]} \BibitemShut {NoStop}%
\bibitem [{\citenamefont {{Abbott}}\ \emph {et~al.}(2018)\citenamefont
  {{Abbott}}, \citenamefont {{Abdalla}}, \citenamefont {{Alarcon}},
  \citenamefont {{Aleksi{\'c}}}, \citenamefont {{Allam}}, \citenamefont
  {{Allen}}, \citenamefont {{Amara}}, \citenamefont {{Annis}}, \citenamefont
  {{Asorey}}, \citenamefont {{Avila}},\ and\ \citenamefont
  {et~al.}}]{1708.01530}%
  \BibitemOpen
  \bibfield  {author} {\bibinfo {author} {\bibfnamefont {T.~M.~C.}\
  \bibnamefont {{Abbott}}}, \bibinfo {author} {\bibfnamefont {F.~B.}\
  \bibnamefont {{Abdalla}}}, \bibinfo {author} {\bibfnamefont {A.}~\bibnamefont
  {{Alarcon}}}, \bibinfo {author} {\bibfnamefont {J.}~\bibnamefont
  {{Aleksi{\'c}}}}, \bibinfo {author} {\bibfnamefont {S.}~\bibnamefont
  {{Allam}}}, \bibinfo {author} {\bibfnamefont {S.}~\bibnamefont {{Allen}}},
  \bibinfo {author} {\bibfnamefont {A.}~\bibnamefont {{Amara}}}, \bibinfo
  {author} {\bibfnamefont {J.}~\bibnamefont {{Annis}}}, \bibinfo {author}
  {\bibfnamefont {J.}~\bibnamefont {{Asorey}}}, \bibinfo {author}
  {\bibfnamefont {S.}~\bibnamefont {{Avila}}},\ and\ \bibinfo {author}
  {\bibnamefont {et~al.}},\ }\bibfield  {title} {\bibinfo {title} {{Dark Energy
  Survey year 1 results: Cosmological constraints from galaxy clustering and
  weak lensing}},\ }\href {https://doi.org/10.1103/PhysRevD.98.043526}
  {\bibfield  {journal} {\bibinfo  {journal} {\prd}\ }\textbf {\bibinfo
  {volume} {98}},\ \bibinfo {eid} {043526} (\bibinfo {year} {2018})},\ \Eprint
  {https://arxiv.org/abs/1708.01530} {arXiv:1708.01530 [astro-ph.CO]}
  \BibitemShut {NoStop}%
\bibitem [{\citenamefont {{Hikage}}\ \emph {et~al.}(2019)\citenamefont
  {{Hikage}}, \citenamefont {{Oguri}}, \citenamefont {{Hamana}}, \citenamefont
  {{More}}, \citenamefont {{Mandelbaum}}, \citenamefont {{Takada}},
  \citenamefont {{K{\"o}hlinger}}, \citenamefont {{Miyatake}}, \citenamefont
  {{Nishizawa}}, \citenamefont {{Aihara}}, \citenamefont {{Armstrong}},
  \citenamefont {{Bosch}}, \citenamefont {{Coupon}}, \citenamefont {{Ducout}},
  \citenamefont {{Ho}}, \citenamefont {{Hsieh}}, \citenamefont {{Komiyama}},
  \citenamefont {{Lanusse}}, \citenamefont {{Leauthaud}}, \citenamefont
  {{Lupton}}, \citenamefont {{Medezinski}}, \citenamefont {{Mineo}},
  \citenamefont {{Miyama}}, \citenamefont {{Miyazaki}}, \citenamefont
  {{Murata}}, \citenamefont {{Murayama}}, \citenamefont {{Shirasaki}},
  \citenamefont {{Sif{\'o}n}}, \citenamefont {{Simet}}, \citenamefont
  {{Speagle}}, \citenamefont {{Spergel}}, \citenamefont {{Strauss}},
  \citenamefont {{Sugiyama}}, \citenamefont {{Tanaka}}, \citenamefont
  {{Utsumi}}, \citenamefont {{Wang}},\ and\ \citenamefont
  {{Yamada}}}]{1809.09148}%
  \BibitemOpen
  \bibfield  {author} {\bibinfo {author} {\bibfnamefont {C.}~\bibnamefont
  {{Hikage}}}, \bibinfo {author} {\bibfnamefont {M.}~\bibnamefont {{Oguri}}},
  \bibinfo {author} {\bibfnamefont {T.}~\bibnamefont {{Hamana}}}, \bibinfo
  {author} {\bibfnamefont {S.}~\bibnamefont {{More}}}, \bibinfo {author}
  {\bibfnamefont {R.}~\bibnamefont {{Mandelbaum}}}, \bibinfo {author}
  {\bibfnamefont {M.}~\bibnamefont {{Takada}}}, \bibinfo {author}
  {\bibfnamefont {F.}~\bibnamefont {{K{\"o}hlinger}}}, \bibinfo {author}
  {\bibfnamefont {H.}~\bibnamefont {{Miyatake}}}, \bibinfo {author}
  {\bibfnamefont {A.~J.}\ \bibnamefont {{Nishizawa}}}, \bibinfo {author}
  {\bibfnamefont {H.}~\bibnamefont {{Aihara}}}, \bibinfo {author}
  {\bibfnamefont {R.}~\bibnamefont {{Armstrong}}}, \bibinfo {author}
  {\bibfnamefont {J.}~\bibnamefont {{Bosch}}}, \bibinfo {author} {\bibfnamefont
  {J.}~\bibnamefont {{Coupon}}}, \bibinfo {author} {\bibfnamefont
  {A.}~\bibnamefont {{Ducout}}}, \bibinfo {author} {\bibfnamefont
  {P.}~\bibnamefont {{Ho}}}, \bibinfo {author} {\bibfnamefont {B.-C.}\
  \bibnamefont {{Hsieh}}}, \bibinfo {author} {\bibfnamefont {Y.}~\bibnamefont
  {{Komiyama}}}, \bibinfo {author} {\bibfnamefont {F.}~\bibnamefont
  {{Lanusse}}}, \bibinfo {author} {\bibfnamefont {A.}~\bibnamefont
  {{Leauthaud}}}, \bibinfo {author} {\bibfnamefont {R.~H.}\ \bibnamefont
  {{Lupton}}}, \bibinfo {author} {\bibfnamefont {E.}~\bibnamefont
  {{Medezinski}}}, \bibinfo {author} {\bibfnamefont {S.}~\bibnamefont
  {{Mineo}}}, \bibinfo {author} {\bibfnamefont {S.}~\bibnamefont {{Miyama}}},
  \bibinfo {author} {\bibfnamefont {S.}~\bibnamefont {{Miyazaki}}}, \bibinfo
  {author} {\bibfnamefont {R.}~\bibnamefont {{Murata}}}, \bibinfo {author}
  {\bibfnamefont {H.}~\bibnamefont {{Murayama}}}, \bibinfo {author}
  {\bibfnamefont {M.}~\bibnamefont {{Shirasaki}}}, \bibinfo {author}
  {\bibfnamefont {C.}~\bibnamefont {{Sif{\'o}n}}}, \bibinfo {author}
  {\bibfnamefont {M.}~\bibnamefont {{Simet}}}, \bibinfo {author} {\bibfnamefont
  {J.}~\bibnamefont {{Speagle}}}, \bibinfo {author} {\bibfnamefont {D.~N.}\
  \bibnamefont {{Spergel}}}, \bibinfo {author} {\bibfnamefont {M.~A.}\
  \bibnamefont {{Strauss}}}, \bibinfo {author} {\bibfnamefont {N.}~\bibnamefont
  {{Sugiyama}}}, \bibinfo {author} {\bibfnamefont {M.}~\bibnamefont
  {{Tanaka}}}, \bibinfo {author} {\bibfnamefont {Y.}~\bibnamefont {{Utsumi}}},
  \bibinfo {author} {\bibfnamefont {S.-Y.}\ \bibnamefont {{Wang}}},\ and\
  \bibinfo {author} {\bibfnamefont {Y.}~\bibnamefont {{Yamada}}},\ }\bibfield
  {title} {\bibinfo {title} {{Cosmology from cosmic shear power spectra with
  Subaru Hyper Suprime-Cam first-year data}},\ }\href
  {https://doi.org/10.1093/pasj/psz010} {\bibfield  {journal} {\bibinfo
  {journal} {\pasj}\ }\textbf {\bibinfo {volume} {71}},\ \bibinfo {eid} {43}
  (\bibinfo {year} {2019})},\ \Eprint {https://arxiv.org/abs/1809.09148}
  {arXiv:1809.09148 [astro-ph.CO]} \BibitemShut {NoStop}%
\bibitem [{\citenamefont {{Refregier}}\ \emph {et~al.}(2010)\citenamefont
  {{Refregier}}, \citenamefont {{Amara}}, \citenamefont {{Kitching}},
  \citenamefont {{Rassat}}, \citenamefont {{Scaramella}},\ and\ \citenamefont
  {{Weller}}}]{1001.0061}%
  \BibitemOpen
  \bibfield  {author} {\bibinfo {author} {\bibfnamefont {A.}~\bibnamefont
  {{Refregier}}}, \bibinfo {author} {\bibfnamefont {A.}~\bibnamefont
  {{Amara}}}, \bibinfo {author} {\bibfnamefont {T.~D.}\ \bibnamefont
  {{Kitching}}}, \bibinfo {author} {\bibfnamefont {A.}~\bibnamefont
  {{Rassat}}}, \bibinfo {author} {\bibfnamefont {R.}~\bibnamefont
  {{Scaramella}}},\ and\ \bibinfo {author} {\bibfnamefont {J.}~\bibnamefont
  {{Weller}}},\ }\bibfield  {title} {\bibinfo {title} {{Euclid Imaging
  Consortium Science Book}},\ }\href@noop {} {\bibfield  {journal} {\bibinfo
  {journal} {arXiv e-prints}\ ,\ \bibinfo {eid} {arXiv:1001.0061}} (\bibinfo
  {year} {2010})},\ \Eprint {https://arxiv.org/abs/1001.0061} {arXiv:1001.0061
  [astro-ph.IM]} \BibitemShut {NoStop}%
\bibitem [{\citenamefont {{Arcelin}}\ \emph {et~al.}(2021)\citenamefont
  {{Arcelin}}, \citenamefont {{Doux}}, \citenamefont {{Aubourg}}, \citenamefont
  {{Roucelle}},\ and\ \citenamefont {{LSST Dark Energy Science
  Collaboration}}}]{2005.12039}%
  \BibitemOpen
  \bibfield  {author} {\bibinfo {author} {\bibfnamefont {B.}~\bibnamefont
  {{Arcelin}}}, \bibinfo {author} {\bibfnamefont {C.}~\bibnamefont {{Doux}}},
  \bibinfo {author} {\bibfnamefont {E.}~\bibnamefont {{Aubourg}}}, \bibinfo
  {author} {\bibfnamefont {C.}~\bibnamefont {{Roucelle}}},\ and\ \bibinfo
  {author} {\bibnamefont {{LSST Dark Energy Science Collaboration}}},\
  }\bibfield  {title} {\bibinfo {title} {{Deblending galaxies with variational
  autoencoders: A joint multiband, multi-instrument approach}},\ }\href
  {https://doi.org/10.1093/mnras/staa3062} {\bibfield  {journal} {\bibinfo
  {journal} {\mnras}\ }\textbf {\bibinfo {volume} {500}},\ \bibinfo {pages}
  {531} (\bibinfo {year} {2021})},\ \Eprint {https://arxiv.org/abs/2005.12039}
  {arXiv:2005.12039 [astro-ph.IM]} \BibitemShut {NoStop}%
\bibitem [{\citenamefont {{Melchior}}\ \emph {et~al.}(2018)\citenamefont
  {{Melchior}}, \citenamefont {{Moolekamp}}, \citenamefont {{Jerdee}},
  \citenamefont {{Armstrong}}, \citenamefont {{Sun}}, \citenamefont {{Bosch}},\
  and\ \citenamefont {{Lupton}}}]{1802.10157}%
  \BibitemOpen
  \bibfield  {author} {\bibinfo {author} {\bibfnamefont {P.}~\bibnamefont
  {{Melchior}}}, \bibinfo {author} {\bibfnamefont {F.}~\bibnamefont
  {{Moolekamp}}}, \bibinfo {author} {\bibfnamefont {M.}~\bibnamefont
  {{Jerdee}}}, \bibinfo {author} {\bibfnamefont {R.}~\bibnamefont
  {{Armstrong}}}, \bibinfo {author} {\bibfnamefont {A.~L.}\ \bibnamefont
  {{Sun}}}, \bibinfo {author} {\bibfnamefont {J.}~\bibnamefont {{Bosch}}},\
  and\ \bibinfo {author} {\bibfnamefont {R.}~\bibnamefont {{Lupton}}},\
  }\bibfield  {title} {\bibinfo {title} {{SCARLET: Source separation in
  multi-band images by Constrained Matrix Factorization}},\ }\href
  {https://doi.org/10.1016/j.ascom.2018.07.001} {\bibfield  {journal} {\bibinfo
   {journal} {Astronomy and Computing}\ }\textbf {\bibinfo {volume} {24}},\
  \bibinfo {eid} {129} (\bibinfo {year} {2018})},\ \Eprint
  {https://arxiv.org/abs/1802.10157} {arXiv:1802.10157 [astro-ph.IM]}
  \BibitemShut {NoStop}%
\bibitem [{\citenamefont {{Lanusse}}\ \emph {et~al.}(2019)\citenamefont
  {{Lanusse}}, \citenamefont {{Melchior}},\ and\ \citenamefont
  {{Moolekamp}}}]{1912.03980}%
  \BibitemOpen
  \bibfield  {author} {\bibinfo {author} {\bibfnamefont {F.}~\bibnamefont
  {{Lanusse}}}, \bibinfo {author} {\bibfnamefont {P.}~\bibnamefont
  {{Melchior}}},\ and\ \bibinfo {author} {\bibfnamefont {F.}~\bibnamefont
  {{Moolekamp}}},\ }\bibfield  {title} {\bibinfo {title} {{Hybrid Physical-Deep
  Learning Model for Astronomical Inverse Problems}},\ }\href@noop {}
  {\bibfield  {journal} {\bibinfo  {journal} {arXiv e-prints}\ ,\ \bibinfo
  {eid} {arXiv:1912.03980}} (\bibinfo {year} {2019})},\ \Eprint
  {https://arxiv.org/abs/1912.03980} {arXiv:1912.03980 [astro-ph.IM]}
  \BibitemShut {NoStop}%
\bibitem [{\citenamefont {Reiman}\ and\ \citenamefont
  {G{\"o}hre}(2019)}]{reiman2019deblending}%
  \BibitemOpen
  \bibfield  {author} {\bibinfo {author} {\bibfnamefont {D.~M.}\ \bibnamefont
  {Reiman}}\ and\ \bibinfo {author} {\bibfnamefont {B.~E.}\ \bibnamefont
  {G{\"o}hre}},\ }\bibfield  {title} {\bibinfo {title} {Deblending galaxy
  superpositions with branched generative adversarial networks},\ }\href@noop
  {} {\bibfield  {journal} {\bibinfo  {journal} {Monthly Notices of the Royal
  Astronomical Society}\ }\textbf {\bibinfo {volume} {485}} (\bibinfo {year}
  {2019})}\BibitemShut {NoStop}%
\bibitem [{\citenamefont {Boucaud}\ \emph {et~al.}(2020)\citenamefont
  {Boucaud}, \citenamefont {Heneka}, \citenamefont {Ishida}, \citenamefont
  {Sedaghat}, \citenamefont {de~Souza}, \citenamefont {Moews}, \citenamefont
  {Dole}, \citenamefont {Castellano}, \citenamefont {Merlin}, \citenamefont
  {Roscani} \emph {et~al.}}]{boucaud2020photometry}%
  \BibitemOpen
  \bibfield  {author} {\bibinfo {author} {\bibfnamefont {A.}~\bibnamefont
  {Boucaud}}, \bibinfo {author} {\bibfnamefont {C.}~\bibnamefont {Heneka}},
  \bibinfo {author} {\bibfnamefont {E.~E.}\ \bibnamefont {Ishida}}, \bibinfo
  {author} {\bibfnamefont {N.}~\bibnamefont {Sedaghat}}, \bibinfo {author}
  {\bibfnamefont {R.~S.}\ \bibnamefont {de~Souza}}, \bibinfo {author}
  {\bibfnamefont {B.}~\bibnamefont {Moews}}, \bibinfo {author} {\bibfnamefont
  {H.}~\bibnamefont {Dole}}, \bibinfo {author} {\bibfnamefont {M.}~\bibnamefont
  {Castellano}}, \bibinfo {author} {\bibfnamefont {E.}~\bibnamefont {Merlin}},
  \bibinfo {author} {\bibfnamefont {V.}~\bibnamefont {Roscani}}, \emph
  {et~al.},\ }\bibfield  {title} {\bibinfo {title} {Photometry of high-redshift
  blended galaxies using deep learning},\ }\href@noop {} {\bibfield  {journal}
  {\bibinfo  {journal} {Monthly Notices of the Royal Astronomical Society}\
  }\textbf {\bibinfo {volume} {491}},\ \bibinfo {pages} {2481} (\bibinfo {year}
  {2020})}\BibitemShut {NoStop}%
\bibitem [{\citenamefont {{Bertin}}\ and\ \citenamefont
  {{Arnouts}}(1996)}]{B&A1996}%
  \BibitemOpen
  \bibfield  {author} {\bibinfo {author} {\bibfnamefont {E.}~\bibnamefont
  {{Bertin}}}\ and\ \bibinfo {author} {\bibfnamefont {S.}~\bibnamefont
  {{Arnouts}}},\ }\bibfield  {title} {\bibinfo {title} {{SExtractor: Software
  for source extraction.}},\ }\href {https://doi.org/10.1051/aas:1996164}
  {\bibfield  {journal} {\bibinfo  {journal} {\aaps}\ }\textbf {\bibinfo
  {volume} {117}},\ \bibinfo {pages} {393} (\bibinfo {year}
  {1996})}\BibitemShut {NoStop}%
\bibitem [{\citenamefont {Zhang}\ \emph {et~al.}(2018)\citenamefont {Zhang},
  \citenamefont {Tian}, \citenamefont {Kong}, \citenamefont {Zhong},\ and\
  \citenamefont {Fu}}]{zhang2018residual}%
  \BibitemOpen
  \bibfield  {author} {\bibinfo {author} {\bibfnamefont {Y.}~\bibnamefont
  {Zhang}}, \bibinfo {author} {\bibfnamefont {Y.}~\bibnamefont {Tian}},
  \bibinfo {author} {\bibfnamefont {Y.}~\bibnamefont {Kong}}, \bibinfo {author}
  {\bibfnamefont {B.}~\bibnamefont {Zhong}},\ and\ \bibinfo {author}
  {\bibfnamefont {Y.}~\bibnamefont {Fu}},\ }\bibfield  {title} {\bibinfo
  {title} {Residual dense network for image super-resolution},\ }in\ \href@noop
  {} {\emph {\bibinfo {booktitle} {Proceedings of the IEEE conference on
  computer vision and pattern recognition}}}\ (\bibinfo {year} {2018})\ pp.\
  \bibinfo {pages} {2472--2481}\BibitemShut {NoStop}%
\bibitem [{\citenamefont {Zhang}\ \emph {et~al.}(2020)\citenamefont {Zhang},
  \citenamefont {Tian}, \citenamefont {Kong}, \citenamefont {Zhong},\ and\
  \citenamefont {Fu}}]{zhang2020residual}%
  \BibitemOpen
  \bibfield  {author} {\bibinfo {author} {\bibfnamefont {Y.}~\bibnamefont
  {Zhang}}, \bibinfo {author} {\bibfnamefont {Y.}~\bibnamefont {Tian}},
  \bibinfo {author} {\bibfnamefont {Y.}~\bibnamefont {Kong}}, \bibinfo {author}
  {\bibfnamefont {B.}~\bibnamefont {Zhong}},\ and\ \bibinfo {author}
  {\bibfnamefont {Y.}~\bibnamefont {Fu}},\ }\bibfield  {title} {\bibinfo
  {title} {Residual dense network for image restoration},\ }\href@noop {}
  {\bibfield  {journal} {\bibinfo  {journal} {IEEE Transactions on Pattern
  Analysis and Machine Intelligence}\ } (\bibinfo {year} {2020})}\BibitemShut
  {NoStop}%
\bibitem [{\citenamefont {Simonyan}\ and\ \citenamefont
  {Zisserman}(2014)}]{simonyan2014very}%
  \BibitemOpen
  \bibfield  {author} {\bibinfo {author} {\bibfnamefont {K.}~\bibnamefont
  {Simonyan}}\ and\ \bibinfo {author} {\bibfnamefont {A.}~\bibnamefont
  {Zisserman}},\ }\bibfield  {title} {\bibinfo {title} {Very deep convolutional
  networks for large-scale image recognition},\ }\href@noop {} {\bibfield
  {journal} {\bibinfo  {journal} {arXiv preprint arXiv:1409.1556}\ } (\bibinfo
  {year} {2014})}\BibitemShut {NoStop}%
\bibitem [{Note1()}]{Note1}%
  \BibitemOpen
  \bibinfo {note} {\protect \url
  {https://github.com/LSSTDESC/BlendingToolKit}}\BibitemShut {NoStop}%
\bibitem [{\citenamefont {{Rowe}}\ \emph {et~al.}(2015)\citenamefont {{Rowe}},
  \citenamefont {{Jarvis}}, \citenamefont {{Mandelbaum}}, \citenamefont
  {{Bernstein}}, \citenamefont {{Bosch}}, \citenamefont {{Simet}},
  \citenamefont {{Meyers}}, \citenamefont {{Kacprzak}}, \citenamefont
  {{Nakajima}}, \citenamefont {{Zuntz}}, \citenamefont {{Miyatake}},
  \citenamefont {{Dietrich}}, \citenamefont {{Armstrong}}, \citenamefont
  {{Melchior}},\ and\ \citenamefont {{Gill}}}]{2015A&C....10..121R}%
  \BibitemOpen
  \bibfield  {author} {\bibinfo {author} {\bibfnamefont {B.~T.~P.}\
  \bibnamefont {{Rowe}}}, \bibinfo {author} {\bibfnamefont {M.}~\bibnamefont
  {{Jarvis}}}, \bibinfo {author} {\bibfnamefont {R.}~\bibnamefont
  {{Mandelbaum}}}, \bibinfo {author} {\bibfnamefont {G.~M.}\ \bibnamefont
  {{Bernstein}}}, \bibinfo {author} {\bibfnamefont {J.}~\bibnamefont
  {{Bosch}}}, \bibinfo {author} {\bibfnamefont {M.}~\bibnamefont {{Simet}}},
  \bibinfo {author} {\bibfnamefont {J.~E.}\ \bibnamefont {{Meyers}}}, \bibinfo
  {author} {\bibfnamefont {T.}~\bibnamefont {{Kacprzak}}}, \bibinfo {author}
  {\bibfnamefont {R.}~\bibnamefont {{Nakajima}}}, \bibinfo {author}
  {\bibfnamefont {J.}~\bibnamefont {{Zuntz}}}, \bibinfo {author} {\bibfnamefont
  {H.}~\bibnamefont {{Miyatake}}}, \bibinfo {author} {\bibfnamefont {J.~P.}\
  \bibnamefont {{Dietrich}}}, \bibinfo {author} {\bibfnamefont
  {R.}~\bibnamefont {{Armstrong}}}, \bibinfo {author} {\bibfnamefont
  {P.}~\bibnamefont {{Melchior}}},\ and\ \bibinfo {author} {\bibfnamefont
  {M.~S.~S.}\ \bibnamefont {{Gill}}},\ }\bibfield  {title} {\bibinfo {title}
  {{GALSIM: The modular galaxy image simulation toolkit}},\ }\href
  {https://doi.org/10.1016/j.ascom.2015.02.002} {\bibfield  {journal} {\bibinfo
   {journal} {Astronomy and Computing}\ }\textbf {\bibinfo {volume} {10}},\
  \bibinfo {pages} {121} (\bibinfo {year} {2015})},\ \Eprint
  {https://arxiv.org/abs/1407.7676} {arXiv:1407.7676 [astro-ph.IM]}
  \BibitemShut {NoStop}%
\bibitem [{Note2()}]{Note2}%
  \BibitemOpen
  \bibinfo {note} {\protect \url
  {https://github.com/LSSTDESC/WeakLensingDeblending}}\BibitemShut {NoStop}%
\bibitem [{\citenamefont {{De Lucia}}\ \emph {et~al.}(2006)\citenamefont {{De
  Lucia}}, \citenamefont {{Springel}}, \citenamefont {{White}}, \citenamefont
  {{Croton}},\ and\ \citenamefont {{Kauffmann}}}]{2006MNRAS.366..499D}%
  \BibitemOpen
  \bibfield  {author} {\bibinfo {author} {\bibfnamefont {G.}~\bibnamefont {{De
  Lucia}}}, \bibinfo {author} {\bibfnamefont {V.}~\bibnamefont {{Springel}}},
  \bibinfo {author} {\bibfnamefont {S.~D.~M.}\ \bibnamefont {{White}}},
  \bibinfo {author} {\bibfnamefont {D.}~\bibnamefont {{Croton}}},\ and\
  \bibinfo {author} {\bibfnamefont {G.}~\bibnamefont {{Kauffmann}}},\
  }\bibfield  {title} {\bibinfo {title} {{The formation history of elliptical
  galaxies}},\ }\href {https://doi.org/10.1111/j.1365-2966.2005.09879.x}
  {\bibfield  {journal} {\bibinfo  {journal} {\mnras}\ }\textbf {\bibinfo
  {volume} {366}},\ \bibinfo {pages} {499} (\bibinfo {year} {2006})},\ \Eprint
  {https://arxiv.org/abs/astro-ph/0509725} {arXiv:astro-ph/0509725 [astro-ph]}
  \BibitemShut {NoStop}%
\bibitem [{\citenamefont {Wang}\ \emph {et~al.}(2004)\citenamefont {Wang},
  \citenamefont {Bovik}, \citenamefont {Sheikh},\ and\ \citenamefont
  {Simoncelli}}]{wang2004image}%
  \BibitemOpen
  \bibfield  {author} {\bibinfo {author} {\bibfnamefont {Z.}~\bibnamefont
  {Wang}}, \bibinfo {author} {\bibfnamefont {A.~C.}\ \bibnamefont {Bovik}},
  \bibinfo {author} {\bibfnamefont {H.~R.}\ \bibnamefont {Sheikh}},\ and\
  \bibinfo {author} {\bibfnamefont {E.~P.}\ \bibnamefont {Simoncelli}},\
  }\bibfield  {title} {\bibinfo {title} {Image quality assessment: from error
  visibility to structural similarity},\ }\href@noop {} {\bibfield  {journal}
  {\bibinfo  {journal} {IEEE transactions on image processing}\ }\textbf
  {\bibinfo {volume} {13}},\ \bibinfo {pages} {600} (\bibinfo {year}
  {2004})}\BibitemShut {NoStop}%
\bibitem [{\citenamefont {{Barbary}}\ \emph {et~al.}(2015)\citenamefont
  {{Barbary}}, \citenamefont {{Boone}},\ and\ \citenamefont
  {{Deil}}}]{sep2015}%
  \BibitemOpen
  \bibfield  {author} {\bibinfo {author} {\bibfnamefont {K.}~\bibnamefont
  {{Barbary}}}, \bibinfo {author} {\bibfnamefont {K.}~\bibnamefont {{Boone}}},\
  and\ \bibinfo {author} {\bibfnamefont {C.}~\bibnamefont {{Deil}}},\ }\href
  {https://doi.org/10.5281/zenodo.15669} {\bibinfo {title} {{Sep: V0.3.0}}}
  (\bibinfo {year} {2015})\BibitemShut {NoStop}%
\bibitem [{Note3()}]{Note3}%
  \BibitemOpen
  \bibinfo {note} {\protect \url
  {https://github.com/esheldon/sxdes}}\BibitemShut {NoStop}%
\end{thebibliography}%


%

\appendix

\section{Iterative results}\label{more_itr}
\begin{figure*}
    \centering
    \includegraphics[width=\linewidth]{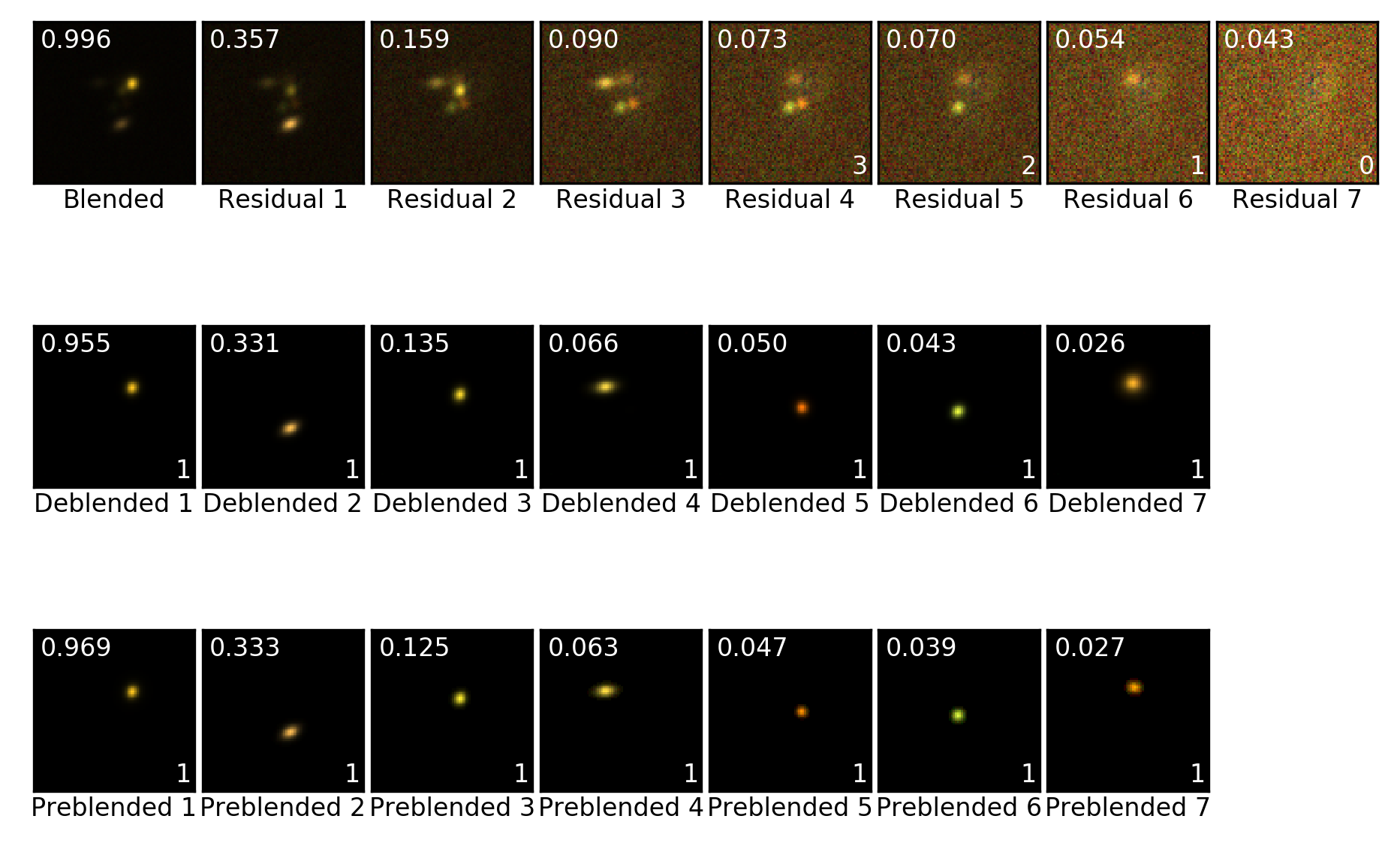}
     \captionsetup{justification=justified}
     \caption{Figure showing iterative results for $7$-galaxy blended images. The classifier was trained for $0$ to $3$-galaxy blended images, therefore the prediction for some of the residual images were skipped.}
    \label{fig:iterative_blend7}
\end{figure*}

FIG. \ref{fig:iterative_blend7} visualizes a typical high-quality deblend for the $7$-galaxy blended images. It shows good recovery of the morphological and color information for all $7$ galaxies. The rest of the galaxies are much fainter compared with the brightest one for the original blended image, but after several iterations, all the residual galaxies emerge for the network to deblend. As the classifier was trained for up to $3$-galaxy blended images, the predictions for images with more than $3$ galaxies are skipped.

PSNR and SSIM results for $4$ to $7$-galaxy blended images are in Table \ref{tab:bigtable_extension}. There are some similar trends as in Table \ref{tab:bigtable}, where the fainter galaxies have higher PSNR and SSIM. This could be interpreted by the normalization of PSNR as in Section \ref{sec:evaluation}.

\begin{table*}
\centering
\begin{tabular}{c|cc|cc|cc|cc|cc|cc|cc|cc}
  \hline \hline
  & \multicolumn{4}{c|}{4-galaxy problem} & \multicolumn{4}{c|}{5-galaxy problem} & \multicolumn{4}{c|}{6-galaxy problem} & \multicolumn{4}{c}{7-galaxy problem}  \\
  \hline
  & \multicolumn{2}{c|}{PSRN} & \multicolumn{2}{c|}{SSIM} & \multicolumn{2}{c|}{PSRN} & \multicolumn{2}{c|}{SSIM} &\multicolumn{2}{c|}{PSRN} & \multicolumn{2}{c|}{SSIM} & \multicolumn{2}{c|}{PSRN} & \multicolumn{2}{c}{SSIM} \\
  \hline
  & Mean  & Median & Mean  & Median & Mean  & Median & Mean  & Median & Mean  & Median & Mean  & Median & Mean  & Median & Mean  & Median \\
\hline 
1 & 53.74  & 56.52   & 0.9946  & 0.9996 & 51.91  & 54.94  & 0.9927  & 0.9994  & 50.57  & 52.90   & 0.9921  & 0.9991 & 49.61  & 51.72  & 0.9907  & 0.9987   \\
2 & 54.17  & 57.82   & 0.9906  & 0.9996 & 52.21  & 55.14  & 0.9879  & 0.9991  & 50.40  & 53.09   & 0.9856  & 0.9983 & 49.60  & 51.00  & 0.9846  & 0.9981   \\
3 & 54.50  & 57.80   & 0.9902  & 0.9994 & 52.22  & 54.35  & 0.9867  & 0.9986  & 50.35  & 50.35   & 0.9843  & 0.9962 & 49.68  & 49.25  & 0.9837  & 0.9947   \\
4 & 56.70  & 59.48   & 0.9948  & 0.9994 & 53.15  & 54.12  & 0.9880  & 0.9977  & 50.67  & 49.96   & 0.9849  & 0.9926 & 50.02  & 48.19  & 0.9847  & 0.9910   \\
5 &        &         &         &        & 56.89  & 58.16  & 0.9934  & 0.9990  & 52.25  & 51.77   & 0.9876  & 0.9940 & 50.52  & 47.72  & 0.9855  & 0.9895   \\
6 &        &         &         &        &        &        &         &         & 56.57  & 56.90   & 0.9924  & 0.9980 & 52.32  & 50.51  & 0.9890  & 0.9923   \\
7 &        &         &         &        &        &        &         &         &        &         &         &        & 56.28  & 55.14  & 0.9925  & 0.9967   \\
  \hline \hline
\end{tabular}
\caption{\label{tab:bigtable_extension}Table showing PSNR(dB) and SSIM for 4 to 7-galaxy deblending problems.}
\end{table*}

\section{RDN recovery} \label{appc}
\begin{figure*} [hbt!]  
    \includegraphics[width=0.75\linewidth]{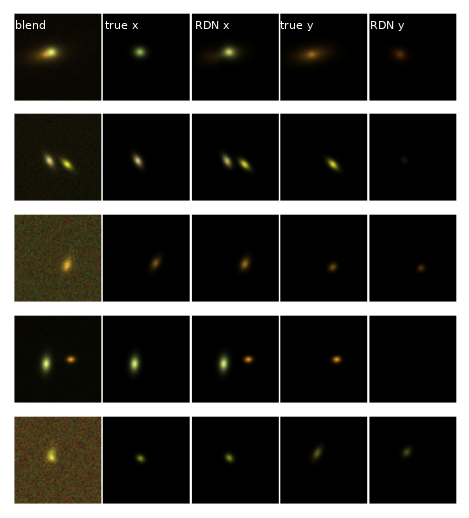}
    \captionsetup{justification=justified}
    \caption{Figure showing some specific failures of the RDN. Row-wise left to right are the blended image, the truth image for the \texttt{x} galaxy, the RDN recovered image for the \texttt{x} galaxy, the truth image for the \texttt{y} galaxy, and the RDN recovered image for the \texttt{y} galaxy. A detailed explanation is given in Appendix \ref{appc}.}
    \label{allfail_1}
\end{figure*}

FIG. \ref{allfail_1} shows a few examples of where the RDN recovered images differ drastically from the truth images. In the first example from the bottom, the true \texttt{x} image is significantly rounder than the RDN recovered \texttt{x} image, while the opposite is true for the respective \texttt{y} images. In the second example, the objects are not deblended at all, as seen in the RDN \texttt{x} image which is identical to the blended image and the extremely faint object in the RDN \texttt{y} image. A possible reason for this is the similarity in the shapes (and to some extent the orientation and color) of both the galaxies in the blended image. In the third case, which is a particularly intertwined blend, the RDN recovered \texttt{x} image is slightly brighter and larger than the truth image, while the \texttt{y} image is the opposite. The fourth case is similar to the second in the sense that both the RDN \texttt{x} and the blended image are identical, but in this case, the RDN \texttt{y} image is blank - however, the two objects in the field do not share a similar shape, orientation or color, and is therefore more mysterious. The fifth example also features a particularly closely associated blend like the third, but in this case, the RDN recovered images for both galaxies is rounder than the truth image.

\end{document}